\documentclass[reprint,amsmath,amssymb,aps
]{revtex4-2}
\usepackage{gensymb}
\usepackage{graphicx}% Include figure files
\usepackage{dcolumn}% Align table columns on decimal point
\usepackage{bm}% bold math
\usepackage{textgreek}
\usepackage{booktabs}
\usepackage{amsmath}
\usepackage{indentfirst}
\usepackage[margin=0.5in]{geometry}
\usepackage{float}
\usepackage{natbib}
\usepackage{xcolor}
\usepackage{caption}

\newcommand{\ArA}[1]{\textcolor{red}{#1}}

\begin{document}
\preprint{APS/123-QED}

\title{Spontaneous brain activity emerges from pairwise interactions in the larval zebrafish brain}

\author{Richard E. Rosch}
\affiliation{Department of Clinical Neurophysiology, King's College Hospital NHS Foundation Trust, London, United Kingdom}
\affiliation{Department of Neurology, Columbia University Irving Medical Center, New York City NY, USA}
\affiliation{Department of Imaging Neuroscience, University College London, London, United Kingdom}
\author{Dominic R. W. Burrows}
\affiliation{MRC Centre for Neurodevelopmental Disorders, King's College London, London, United Kingdom}
\author{Christopher W. Lynn}
\affiliation{Initiative for the Theoretical Sciences, Graduate Center, City University of New York, New York, USA}
\affiliation{Joseph Henry Laboratories of Physics and Lewis–Sigler Institute for Integrative Genomics, Princeton University, Princeton, NJ, USA}
\author{Arian Ashourvan}
\affiliation{Department of Psychology, University of Kansas, Lawrence, USA}
\email{Correspondence to: ashourvan@ku.edu}

\date{December 2022}

\begin{abstract}
Brain activity is characterized by brain-wide spatiotemporal patterns which emerge from synapse-mediated interactions between individual neurons. Calcium imaging provides access to \textit{in vivo} recordings of whole-brain activity at single-neuron resolution and, therefore, allows the study of how large-scale brain dynamics emerge from local activity. In this study, we used a statistical mechanics approach---the pairwise maximum entropy model (MEM)---to infer microscopic network features from collective patterns of activity in the larval zebrafish brain, and relate these features to the emergence of observed whole-brain dynamics. Our findings indicate that the pairwise interactions between neural populations and their intrinsic activity states are sufficient to explain observed whole-brain dynamics. In fact, the pairwise relationships between neuronal populations estimated with the MEM strongly correspond to observed structural connectivity patterns. Model simulations also demonstrated how tuning pairwise neuronal interactions drives transitions between critical and pathologically hyper-excitable whole-brain regimes. Finally, we use virtual resection to identify the brain structures that are important for maintaining the brain in a critical regime. Together, our results indicate that whole-brain activity emerges out of a complex dynamical system that transitions between basins of attraction whose strength and topology depend on the connectivity between brain areas.  

\end{abstract}
\maketitle

\section{\label{sec:level1}Introduction}

Macroscopic dynamics emerge out of interactions between system components at the microscale. In the brain, neuronal action potentials are caused by the interplay between ionic conductances and membrane channel proteins \cite{bean2007action}, while neuronal network dynamics emerge from the synaptic connections formed between constituent neurons and their intrinsic activity \cite{buzsaki2010neural}. At the largest scales, fluctuations in whole-brain dynamics might be driven by a diversity of molecular and neuronal behaviors at fine-grained scales. However, as in many physical systems, much of this microscale complexity will have a negligible influence over the macroscale properties of the system and, therefore, may be ignored when considering macroscopic brain dynamics. Investigating which microscale properties constrain macroscale brain dynamics is fundamental to understanding brain function, as macroscopic network dynamics are strongly linked to behavior and cognition \cite{marek2022frontoparietal,raichle2015brain,seeley2019salience,vossel2014dorsal}.

To bridge the gap between the microscale and macroscale, we require models that can reproduce observed microscopic details while making minimal assumptions about the rest of the system. In other words, we require the maximum entropy model (MEM) consistent with the observed microscopic properties \cite{jaynes1957information}. Here, we seek to understand whether collective patterns of neural activity can be viewed as emerging from the network of pairwise correlations between neuronal populations. To answer this question, we employ the pairwise MEM, which is tuned to match the observed average activities and pairwise correlations in neural activity, but remains explicitly agnostic to higher-order relationships \cite{schneidman2006weak}. From these microscopic details, the resulting model predicts the distribution over macroscopic states of collective neural activity. In this way, the pairwise MEM allows us to understand whether and to what extent macroscopic properties can arise from simple pairwise relationships at the microscale. Furthermore, once a pairwise MEM has been constructed, it can be used to make direct analogies to statistical physics, by constructing an energy landscape that defines the attractor states of the collective neural dynamics \cite{tkavcik2015thermodynamics}.

Several studies have demonstrated the utility of maximum entropy approaches in explaining brain activity in small neuronal populations, recorded from salamander retina \cite{schneidman2006weak,tkavcik2014searching}, rat prefrontal cortex \cite{tavoni2017functional}, and cat parietal cortex \cite{marre2009prediction}. However, it remains unclear whether such minimal-assumption approaches may also be applied to entire brain network dynamics. Recent advancements in human whole-brain neuroimaging and large-scale computational models have identified key relationships between microscale and macroscale dynamics in the brain---indicating roles for local inhibition in shaping macroscopic functional connectivity \cite{deco2014local}, neurotransmitter dynamics in driving emergent whole-brain states \cite{kringelbach2020dynamic}, and neuronal population parameter shifts in causing pathological macroscopic brain states during seizures \cite{jirsa2017virtual,rosch2018calcium}. However, data derived from such whole-brain imaging approaches are typically recorded at poor spatial resolutions, such that neuronal activity is coarse-grained into units consisting of millions of neurons, and such recordings are insensitive to the heterogeneity of microscale activity \cite{kirschstein2009source}. Therefore, to accurately test whether MEM is a valid model of macroscopic brain dynamics, we require techniques that enable whole-brain recordings with microscopic resolution. Here, we take advantage of the larval zebrafish, which has an optically accessible nervous system enabling the recording of single-neuron activity across the entire brain \cite{ahrens2013whole}, making it ideal for the interrogation of multi-scale brain dynamics \cite{burrows2020imaging}. Interestingly, microscale neuronal connectivity has been shown to regulate global brain states \cite{van2023neural,ponce2018whole} (19,20), and when dysregulated, can drive macroscopic brain dysfunction \cite{burrows2023microscale}. Therefore, probing micro-macroscale relationships through MEMs using the larval zebrafish can provide mechanistic insight into the origins of brain-wide dynamics in health and disease. 

Here, we estimate the pairwise interactions between neural populations at different scales to understand the role of microscopic neuronal dynamics in shaping functional and pathological macroscopic brain states. To do this, we studied resting state whole brain dynamics recorded from 11 zebrafish larvae ($\approx$ 84 minutes of concatenated time series) \cite{chen2018brain}. We used spatial and functional clustering to identify functional populations of neurons at different scales whose pairwise interactions best explained macroscopic brain activity. Our results demonstrate that spontaneous patterns of whole brain activity at larger scales (i.e., 12 regions) can be modeled as a stochastic process dictated only by cluster activation propensity and inter-cluster connectivity patterns. Moreover, we show that the estimated pairwise (i.e., second-order) model parameters echo the structural connectivity that has been reported between macroscopic brain areas. By reconstructing the basins of attraction between different states in the model, we can explain the state transition patterns observed in macroscopic brain data, thus demonstrating how low-order interactions between coarse clusters drive macroscopic brain dynamics. Finally, we present a numerical simulation to demonstrate that tuning the coupling between clusters through a ferromagnetic phase transition drives the dynamics from a critical to a hyper-synchronous regime, resembling the transition to pathological hypersynchronous activity seen, for example, in epileptic seizures, thus predicting emergent brain dysfunction from pairwise coupling changes at lower scales. Our results suggest that a simple brain model based on the probabilistic interactions between regions governed by the strength of the connectivity patterns can mechanistically explain and predict emergent functional and pathological activity across the whole brain.

\section{\label{sec:level1}Results}
\subsection{\label{sec:level2} Dimensionality reduction and functional neuronal clusters}

Dimensionality reduction on high-dimensional neuroimaging datasets is commonly performed before fitting to a pairwise maximum entropy model for several reasons (e.g., \cite{schneidman2006weak, watanabe2013pairwise,ashourvan2017energy}). First, fitting MEMs to high-dimensional datasets can be computationally expensive, making dimensionality reduction a necessary first step. Second, dimensionality reduction helps to identify and preserve the most relevant features in the dataset, uncovering patterns and relationships in the data that might not be immediately apparent in high-dimensional spaces. In this way, dimensionality reduction can provide insights into the relevant scales in the system's dynamics, leading to more interpretable models and results.

\begin{figure*}%[hbtp]
\centering
\includegraphics[width=1\linewidth ]{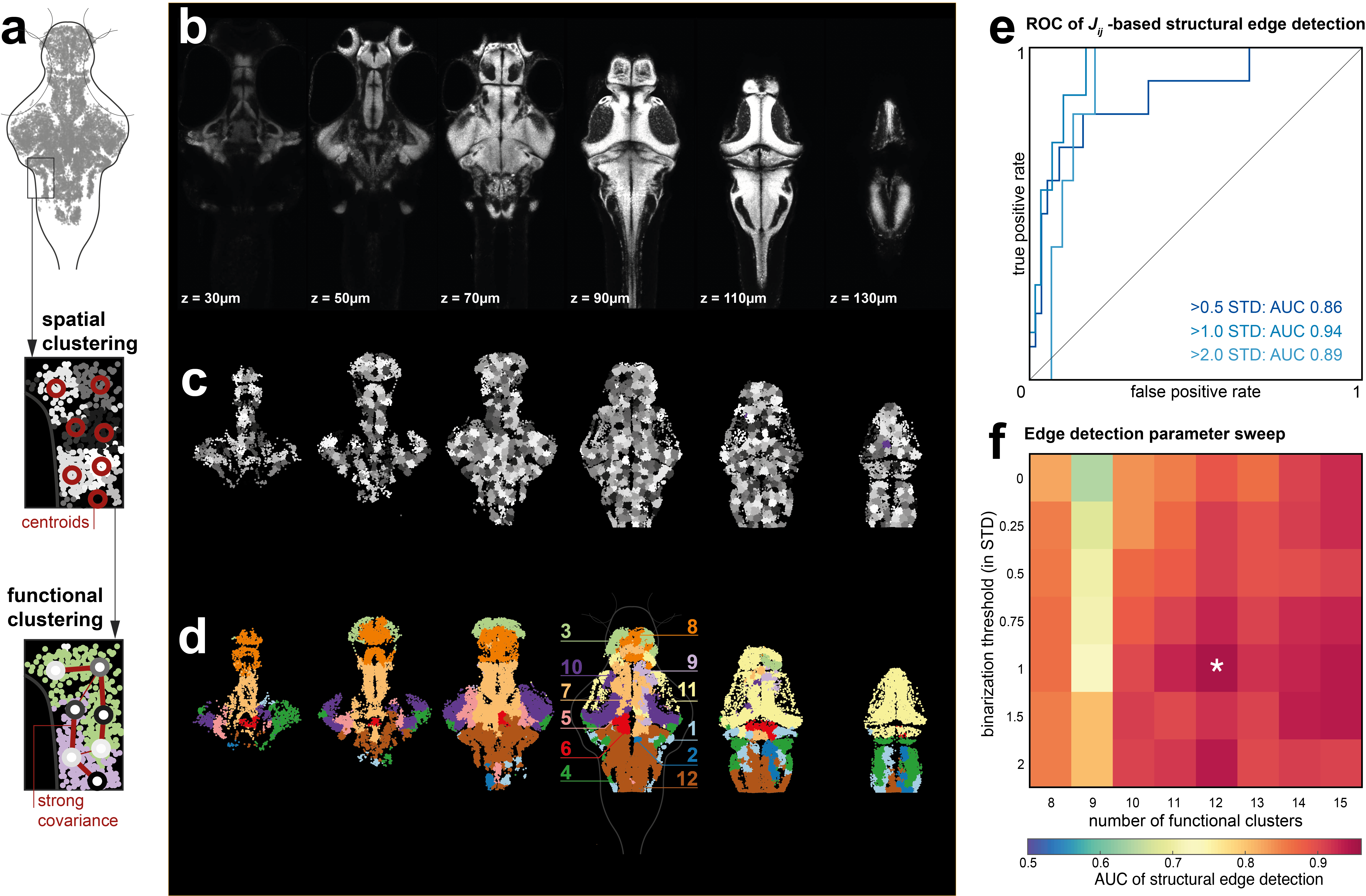}
\caption{\textbf{Coarse-grained functional clustering of individual neurons recorded across the zebrafish brain} \textbf{a.} Workflow illustration--(i, top) neuron positions across multiple fish are coregistered to standard space, (ii, middle) neurons are allocated to 1000 $k$-means clusters based on $xyz$-cell position in standard space, (iii, bottom) spatial clusters of neurons are assigned a smaller number of functional clusters based on clustering of their covariance. \textbf{b.} Horizontal sections through the zebrafish brain in the standard atlas space, \textbf{c.} Spatial cluster assignments of individual neurons, \textbf{d.} Functional cluster assignment of individual neurons. \textbf{e.} AUC values for detection of structural edges based on $J_{ij}$ matrix estimated based on clusters of different sizes and at different binarization thresholds of average cluster activations. The white * marker shows the highest AUC value. \textbf{f.} ROC for detecting structural edges (i.e., fiber count between clusters) for different structural edge thresholds (defined in relation to the standard deviation of all edges). \textbf{g.} Detail of single z-slice demonstrating the spatial location of 12 functional clusters in relation to the major anterior-posterior organization of the larval zebrafish brain.}
\label{figure_1}
\end{figure*}

We identify the neurons that are functionally correlated and, therefore, may belong to a cluster via a two-step process. First, we realign all brains to the larval zebrafish Z-brain template atlas \cite{randlett2015whole} and concatenate neurons from all larvae in three-dimensional space. Next, we used $k-$means clustering to identify 1000 anatomical regions-of-interest (ROI) based on neurons' spatial proximity (Fig. \ref{figure_1}a). Second, we used the covariance matrix of anatomical ROIs' average calcium traces to identify the hierarchical functional clusters at different resolutions. We present the identified large-scale functional clusters for $N= 12$ in Fig. \ref{figure_1}d, and the number of cells per cluster distributions in SI Fig. 1. Additional clustering analyses reveal an optimal number of clusters at $N=2$ and no clear optimal number at higher resolutions (for more details, see SI Fig. 2).        

\subsection{Pairwise MEM interaction weights echo the underlying structural connectivity}

As outlined above, a pairwise MEM is defined solely by the average activities of different brain regions and their functional correlations. From these microscopic details, the pairwise MEM then makes the maximally unbiased predictions for the probabilities of collective states of activity of the entire system without any additional constraints. Specifically, we treat the activity of each region as a binary variable, either `on' ($\sigma_{i} = +1 $) or `off' ($\sigma_{i} = -1$), such that the collective activity of all regions is defined by the vector $\sigma$. We seek the distribution $P(\sigma)$ over collective activity patterns that is consistent with the observed individual activation rates $\left \langle \sigma_{i} \right \rangle$ and the pairwise correlations $\left \langle \sigma_{i}\sigma_{j} \right \rangle$, but otherwise has maximum entropy. This is the pairwise MEM:

\begin{equation}
    P(\sigma)=\frac{1}{Z}\exp\left [ \sum_{i=1}^{N} h_{i}\sigma_{i} +\frac{1}{2}\sum_{i\neq j}^{N} J_{ij}\sigma_{i}\sigma_{j} \right ],
\label{eq_1}
\end{equation}

\noindent where $Z$ is the normalizing partition function, and the parameters $h_{i}$ and $J_{ij}$ represent the activation biases of individual regions and the pairwise couplings between regions, respectively (see Materials and methods for more details). Importantly, the parameters $h_i$ and $J_{ij}$ are tuned such that the model matches the microscopic statistics $\left \langle \sigma_{i} \right \rangle$ and $\left \langle \sigma_{i}\sigma_{j} \right \rangle$ observed in the real system.

We hypothesize that the estimated $J_{ij}$, which capture the functional interactions between regions in the model, would show high similarity to the anatomical axonal (i.e., fiber count) connectivity between regions. As anticipated, a significant correlation was discovered between the estimated $J_{ij}$ values and the strengths of structural connections (SI Fig. 3). Likewise, the strength of the estimated functional interactions $J_{ij}$ could reliably predict the presence of anatomical connectivity (i.e., binarized structural connectivity) between clusters, as indicated by high area under the curve (AUC) values of the Receiver Operating Characteristic (ROC) Curves (Fig. \ref{figure_1}e). We utilise the AUC of the ROC as a measure for the performance of binary classification of structural connectivity edges as present or absence based on $J_{ij}$ edge weights across a range of binarization thresholds. Here we show the binarization threshold ($30\%$ of maximum fiber count between clusters) that leads to the highest AUC values. We present similar results with different binarization thresholds in SI Fig. 4.

We can leverage this relationship between the inferred functional interactions $J_{ij}$ and ground-truth structural connectivity to find the topological scale (i.e., clustering resolution) that maximizes this structure-function coupling. Although the small number of clusters (e.g., 5-7) yield the highest similarity between the structural connectivity and $J_{ij}$ matrices, several clusters are spatially disjointed and scattered in both the anterior and posterior regions of the brain (SI Fig. 3e). The next topological scale with high correlations (SI Fig. 3) and AUC values (Fig. \ref{figure_1}e) is at $N=12$ clusters (at the 'on' state threshold of $z=1$ for z-scored cluster activation time series), which will be used for the remainder of the analyses presented here. We also found converging results (SI Fig. 3 and SI Fig. 5) for larger numbers of clusters ($N>15$) with the MEM estimated using the pseudo-likelihood maximization method (see Materials and Methods for more details). 

In contrast to clustering with fewer, but more distributed individual clusters, the $N=12$ resolution does not exhibit anterior/posterior scattered clusters. Instead, the identified clusters are local and largely contiguous, as demonstrated in SI Fig 3e. Based on our hypothesis that the long-distance functional relationships between neuronal populations rely on the underlying anatomical connections between regions, we investigate the system's behavior at $N=12$ clusters (as depicted in Fig. \ref{figure_1}d). This topological scale maximizes the structure/function coupling and results in clusters characterized by local, contiguous populations scattered across the brain.                

\begin{figure}%[hbtp]
\centering
\includegraphics[width=1\linewidth ]{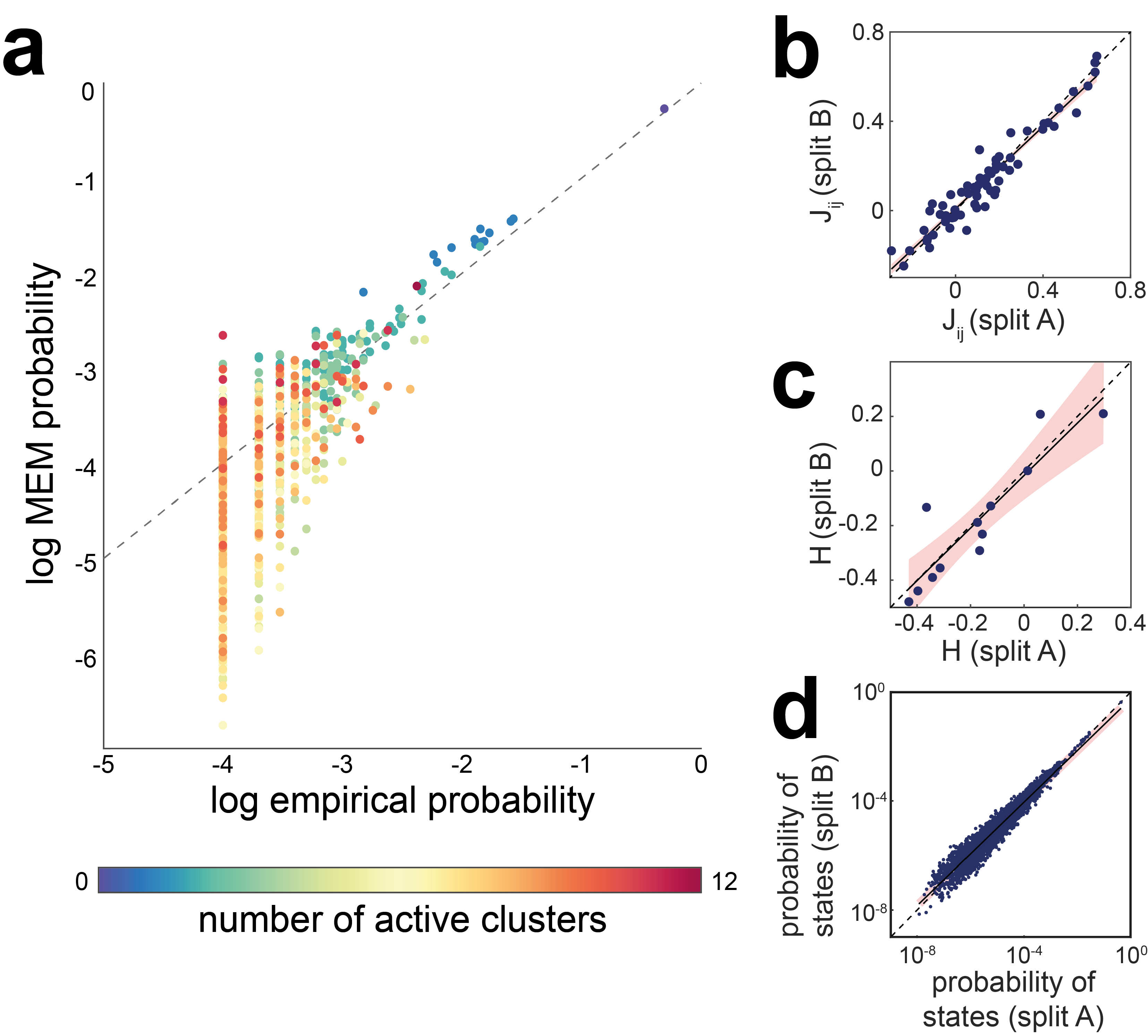}
\caption{\textbf{MEM robustly predicts whole-brain activation patterns} \textbf{a.} Pairwise MEM-estimated and empirical probability of all states. Colors represent the number of active clusters. \textbf{b-d.} Robustness of fit against noise within the data. Full datasets are randomly split into equally sized samples $A$ and $B$. Correlation of sample $A$ and $B$ estimates of \textbf{b} $J_{ij}$, \textbf{c} $h$, and \textbf{c} the probabilities of all states are shown.} 
\label{figure_2}
\end{figure}

\subsection{The pairwise MEM accurately predicts the probabilities of large-scale activation patterns}

We hypothesize that the spontaneous activity of large-scale networks can be described as a stochastic process constrained by the activation rates of individual neuronal clusters and the strength of the functional connectivity between them. To test this hypothesis, we fit the pairwise MEM to the binarized patterns of collective cluster activity. See the Materials and Methods section for details regarding the pairwise MEM model and parameter estimation. Our results demonstrate that the pairwise MEM accurately predicts the probabilities of commonly observed states (Fig. \ref{figure_2}a). Prediction accuracy is overall lower for states with low probability, but is robust against inherent noise in the data (Fig \ref{figure_2}b-d).

We observe that the pairwise MEM offers a statistically supported model of observed activity, as it accounts for $\approx 84 \%$ of the multi-information that the first-order (i.e., independent) model does not capture. In SI Fig. 6, we show the goodness-of-fit of the model monotonically decreases by increasing the number of clusters. Increasing the activation binarization threshold similarly lowers the goodness-of-fit at the smaller number of clusters. However, this relationship gradually reverses towards larger numbers of clusters ($N>13$). In the Materials and Methods, we discuss the model's goodness-of-fit in detail. Together these results suggest that the Pairwise MEM accurately describes the probability of the spontaneous large-scale activation patterns in the zebrafish brain.

\subsection{Stochastic transitions between attractor basins of neural states}

\begin{figure}%[hbtp]
\centering
\includegraphics[width=1\linewidth ]{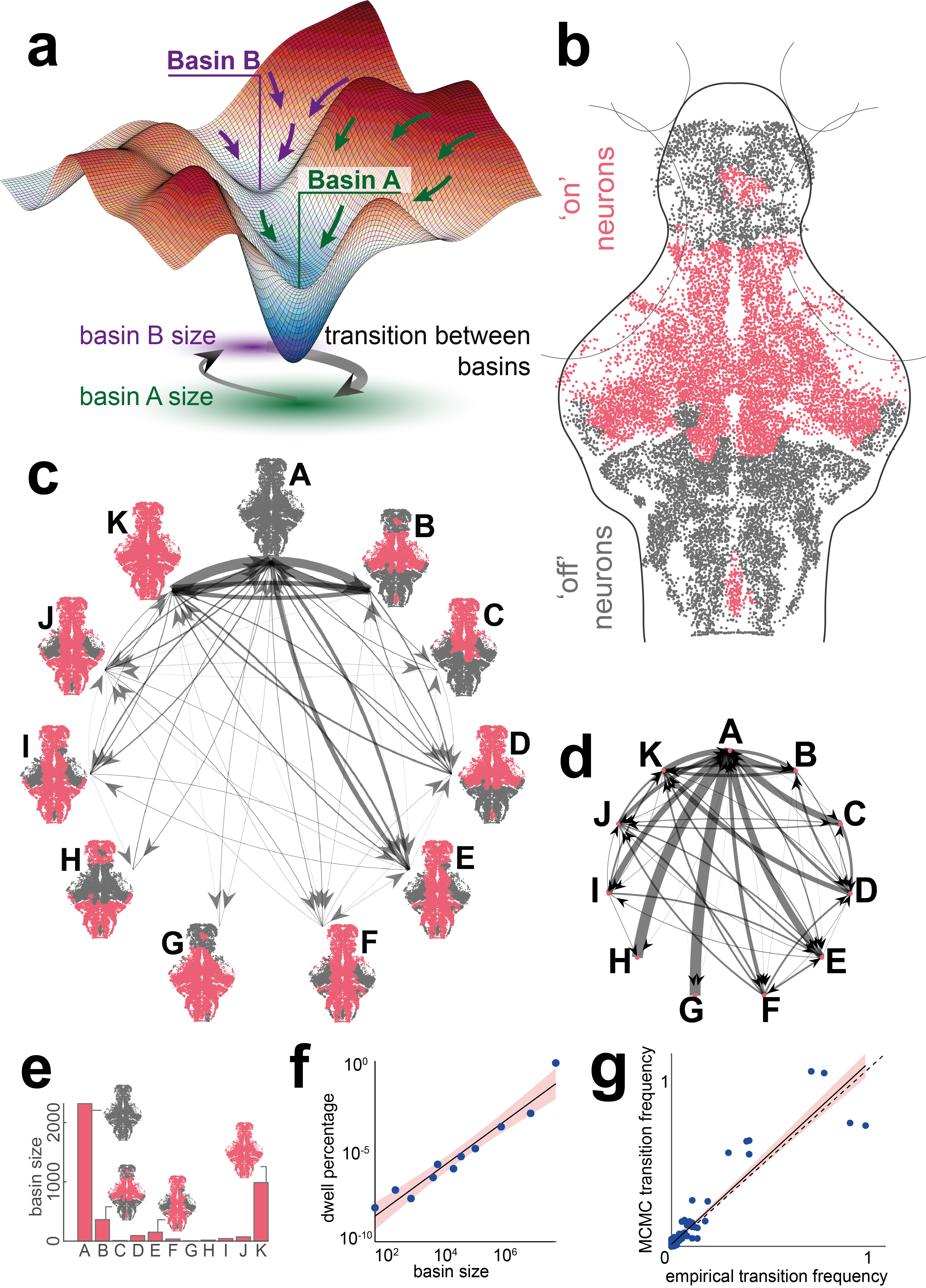}
\caption{\textbf{Attractor basins shape the energy landscape of whole-brain activation patterns} \textbf{a.} Cartoon representation of high-dimensional energy landscape in which attractor states are defined as local minima, with surrounding basins consisting of brain states with transitions tending towards the attractor state. \textbf{b.} Representative attractor state ('B') consisting of on/off patterns across neuron clusters. \textbf{c.} Empirically observed transition frequencies  between 11 attractor states ($A$-$K$) identified for $N=12$ clusters, with heterogeneous and asymmetric observed transition frequencies, with most frequent transitions from and two the 'all-off' state $A$; thicker lines indicate a higher absolute number of transitions observed.  \textbf{d.} Empirical transition probabilities between the 11 attractor states in panel \textbf{c}; thicker lines indicate higher transition probabilities from any one state, with the sum of all outgoing arrows being equal to 1. \textbf{e.} Basin size for 11 attractor states. \textbf{f.} Relationship between observed dwell times and the basin size of each attractor state. \textbf{g} Relationship between empirically observed transition frequencies and MCMC predicted transition frequencies based on the energy landscape. }

\label{figure_3}
\end{figure}

Next, we set out to understand the constraints that define transitions between macroscopic brain states, particularly to understand if such transitions may be driven by a stochastic process over the energy landscape as defined by the MEM. Now that we have confirmed that the pairwise MEM accurately describes the large-scale neural patterns, we can use the model to explore the energy landscape over neural states. We first identified the attractor states of the estimated model by constructing a multi-dimensional mesh of all states, which we refer to as the energy landscape. In this landscape, neighboring activation states only differ in the activity of a single cluster. We exhaustively searched the energy landscape using a steep search algorithm to identify the states that are local maxima of the probability distribution (i.e., local minima of the energy landscape). The local minima were defined as states with higher probabilities than their neighboring states. To identify the states that belong to the basin of each local minimum, we found all neighboring states that transition on a downward gradient towards the minimum (Fig. \ref{figure_3}a-b; see Materials and Methods).

In Fig. \ref{figure_3}c, we show the identified local minima for the $N=12$ functional clusters and the empirical transition frequency between their basins. Our results also show an exponential relationship between the size (the number of basin states) and the dwell time of local minima basins (Fig. \ref{figure_3}f). These results demonstrate the significant influence of high-probability local minima with large basins of attraction. To determine if a stochastic process can explain the patterns of state transitions over the estimated energy landscape, we performed a random walk process using the Markov chain Monte Carlo (MCMC) algorithm (see Materials and Methods). The results of the MCMC simulations showed that in large-scale clusters, the frequency of transitions between the basins of the local minima closely matched the empirical transition frequencies (Fig. \ref{figure_3}g). This relationship was statistically significant, with a linear regression \ArA{$p$-value of $4.6 \times 10^{-70}$ and an $R^{2}$ value of 0.91.} This indicates that macroscopic state transitions are constrained by the topology of the energy landscape, defined by the relative probabilities of neighboring brain states. 

To better understand how the topology of the energy landscape contributes to the observed state transitions, we also examined features of the attractor states and their basins, such as the energy barrier between the local minima and the basin's size. Specifically, we first constructed the disconnectivity graph that reveals the relationships between the local minima across different energy levels (SI Fig. 7). The energy barrier between two local minima is the difference between the local minima and saddle states' energy (i.e., the lowest energy level connecting two local minima). Our results demonstrate that the energy barriers between minima show a weak relationship with the frequency of transitions between local minima basins, but the linear fit is poor and the overall correlation is low (SI Fig. 8).

\subsection{Seizure-like transitions from disorder to order}

Next, we utilized the MEM to study transitions between physiological and pathological whole-brain states. For example, epileptic seizures are characterized by paroxysmal transitions of macroscopic brain dynamics into an abnormal dynamical regime \cite{stafstrom2015seizures}. These transitions are a common feature of brain pathophysiology across species \cite{jirsa2014nature,podell1996seizures}, and represent a huge burden for patients \cite{cardarelli2010burden}. To study transitions between distinct dynamical states, such as physiological brain activity and hypersynchronous activity seen during epileptic seizures, we took advantage of the characteristic phase transition in the pairwise MEM that occurs between large-scale ordered and disordered states. Such phase transitions are best described by the Ising model, which is mathematically identical to the pairwise MEM. This model was developed in statistical mechanics to explain the transition between ferromagnetic and paramagnetic states in magnetic materials by modeling the interaction of atomic spins. Here, parallel spins have lower energy than anti-parallel spins, driving the tendency for atoms to share the same spin, giving rise to large-scale order and ferromagnetism. However, if the temperature in the system is increased above the critical temperature, $T_{c}$, thermal fluctuations disrupt the neighboring spin correlations, causing a disordered mixture of parallel and anti-parallel spins, the paramagnetic phase. This sudden shift from ferromagnetic to paramagnetic behavior with only small changes around $T_{c}$ is known as a \emph{phase transition}.

\begin{figure*}%[hbtp]
\centering
\includegraphics[width=0.85\linewidth ]{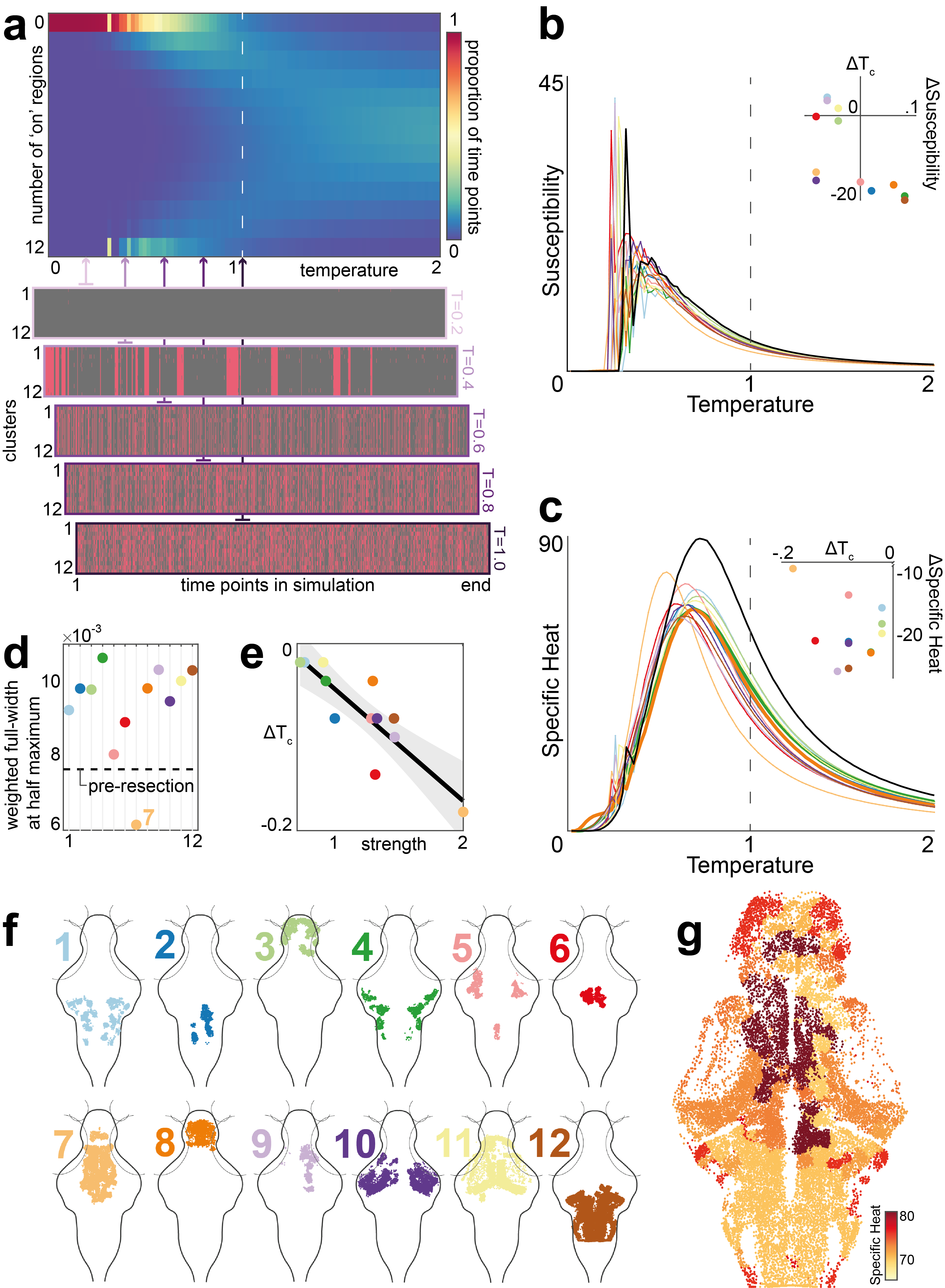}
\caption{\textbf{Seizure-like phase transitions of global activity in the pairwise MEM.} \textbf{a.} (top) Proportion of simulated states with different numbers of active clusters at different temperatures, simulated using the MCMC algorithm. (bottom) The simulated state transitions at five different sample temperatures. Pink indicates clusters in the 'on' state; grey indicates clusters in the 'off' state. \textbf{b-c.} The susceptibility (\textbf{b}) and the specific heat (\textbf{c}) curves before (black) and after (color-coded for each cluster) virtual resection. The insets show the change in the peak of the curves following the resections. \textbf{d.} Full width at half maximum (FWHM) values of the specific heat curves. \textbf{e.} Post-resection change in the critical temperature, $T_{c}$, and the strength of the clusters' functional connections (i.e., the sum of $J_{ij}$ rows). \textbf{f.} Individual functional clusters, projected across all $z$-layers onto a the $xy$ zebrafish brain outline. \textbf{g.} Post-resection values of specific heat maximum mapped onto individual neuron positions, plotted for a single $z$-layer } 
\label{figure_4}
\end{figure*}

Here we leverage the same principles to model transitions from physiological to pathological macroscopic brain dynamics and to probe the microscopic interactions from which they arise. Seizure transitions can be induced by globally increasing the excitability of brain networks, for example, through drugs blocking the inhibitory neurotransmitter GABA or enhancing excitatory transmission through glutamate \cite{Levesque2016chemoconvulsant, shimada2018pentylenetetrazole}. In the more abstract formulation of pairwise MEM, we represent changes in effective synaptic connectivity through changes in the temperature parameter. In the pairwise MEM (equation \ref{eq_1}), the temperature is usually set to $T=1$. However, by dividing the exponent in equation \ref{eq_1} by $T$ (see Materials and Methods), the Ising model can simulate and predict the effects of changing temperature on the large-scale behavior of the system. Here, we modeled a global increase in excitability by a decrease in the temperature of the MEM (Fig. \ref{figure_4}). Mathematically, as the temperature increases, the probability distribution becomes more spread out, allowing for a broader range of energy states to be occupied. This increased spread in energy states corresponds to increased randomness in the system. By performing MCMC simulations under varying values of $T$ (binarization threshold $z=0$, see SI Fig. 9 for $z=1$), we demonstrate that reducing the system's temperature leads to a transition in the global order of the neural activity (Fig. \ref{figure_4}a). Namely, the system transitions from disordered activation patterns (i.e., zero mean activation) at higher temperatures ($T > 1$) to highly ordered, hypersynchronous seizure-like states, marked by periods of brain-wide activation followed by brain-wide inactivity analogous to a post-seizure state\cite{Bruno_Richardson_2020} (e.g., Fig. \ref{figure_4}a, $T=0.4$). This demonstrates that the empirically-derived MEM may be used to model pathological whole-brain state transitions. Next, we aimed to perturb empirically-derived MEM parameters to ascertain the contribution of observed biological features in driving physiological-pathological state transitions.

\subsection{Regional contribution to critical dynamics}

Since the intrinsic connectivity of brain structures is heterogeneous, each region shapes global dynamics differently. For instance, one could hypothesize that densely connected brain regions with widely distributed connections play an important role in maintaining the system's dynamics poised between order and disorder, a regime known as criticality. Interestingly, the critical regime may be a favourable regime for brain dynamics \cite{Beggs2003, Shew2009}, and brain pathology might emerge as a loss of criticality \cite{Zimmern2020}. Therefore, studying how different brain structures regulate transitions about the critical regime can help us to understand how neuronal populations drive optimal and pathological whole-brain regimes. The organizational role of different brain regions can be examined empirically or in silico by removing or ablating single regions and their connections, and assessing changes in the macroscopic behavior of the system. Here, we examine the effect of removing individual regions on the transition between order and disorder in the MEM.

To investigate the transition between order and disorder, we study the specific heat of the system, which measures the change in the average energy due to a slight change in temperature (see Materials and Methods). The temperature associated with the peak specific heat (Fig. \ref{figure_4}a) can be used to identify the critical temperature $T_{c}$ of a given system. When a region is removed from the model, it no longer influences the rest of the system. This can significantly affect specific heat (Fig. \ref{figure_4}b-c, depending on the nature of the interactions with the removed region. For instance, if the interactions with the removed region are strong, removing them effectively reduces global connectivity, which is equivalent to an apparent increase in the temperature of the system, shifting the specific heat curve to the left (Fig. \ref{figure_4}c) -- resulting in critical transitions being achieved at lower temperatures. Fig. \ref{figure_4}c shows changes in the specific heat curves (and, in turn, the critical temperature $T_c$) induced by removing different regions. A decrease in the full width at half maximum (FWHM) of the specific heat curve would correspond to a \emph{sharpening} of the specific heat curves after resection (Fig. \ref{figure_4}d). This is seen only when removing cluster 7, suggesting that this intervention makes the transition \emph{sharper}, with a smaller temperature window for critical behavior. Cluster 7 corresponds to the diencephalon which contains the thalamus and hypothalamus among other regions, with known roles for homeostatic regulation and sensori-motor integration. This finding indicates that the connectivity of the diencephalon supports a broad range of network states to reside in the vicinity of the critical regime. 

Interestingly, our results show that the specific heat decreases and peaks at lower temperatures for all resections, compared with the full system. Therefore, the connections provided by each region tend to push the system closer to criticality when compared to a system lacking this region, but with all other parameters being equal (e.g. the same connection weights between all other regions). However, if the interactions with the removed region are weak (e.g., clusters 1 and 3), removing them might only have a negligible effect on the system, resulting in a minor change on the specific heat. In fact, we observed a close relationship between the overall strength of a cluster's functional interactions and its post-resection effect on the critical temperature (Fig. \ref{figure_4}e).   

Removing a region from the model can also affect other system properties, such as magnetization and susceptibility. The susceptibility measures the change in the net magnetization (i.e., average activation across all regions) due to small changes in the external field, $h_{i}$, which roughly corresponds to regional excitability or activation propensity (see Materials and Methods). The post-resection susceptibility measures echo the specific heat results (Fig. \ref{figure_4}b) and identify similar region-specific effects. Finally, we also explored the effect of increasing the cluster activation binarization threshold on the above-mentioned results. SI Figure 10 shows that the resection of some of the similar clusters, such as cluster 7, also results in the biggest change in the susceptibility and specific heat curves at higher binarization thresholds (e.g., $z=1$). Taken together, these findings suggest regional specificity in the regulation of whole-brain critical dynamics, with hub-like regions exhibiting particularly strong control over the critical regime. 

\section{Discussion}
\subsection{Capturing whole brain dynamics in the larval zebrafish brain through minimal models of pairwise interactions}

Neuronal systems display complex population dynamics that unfold as observable patterns of whole-brain activity. Whether these patterns emerge from simple pairwise interactions between brain regions remains a central open question. The pairwise MEM approach presented here accounts for the observed covariance (i.e., functional connectivity) between regions, but remains explicitly agnostic to all higher--order interactions. The MEM includes two key terms: intrinsic neuronal activity ($h_i$) and pairwise interactions ($J_{ij}$). Here, we show that the probabilities of collective brain states can be described accurately by the pairwise MEM, indicating that macroscopic features of neural activity can arising from fine-scale interactions.

Tested in an empirical dataset with single-neuron resolution, previous work in zebrafish had already demonstrated that simple pairwise interactions between neuronal ensembles could reproduce statistical characteristics of whole-brain dynamics \cite{van2023neural}. In the dataset presented here, we aimed to predict the probabilities of specific whole-brain states at coarse-grained scales. In finer-scale, higher-dimensional data, the pairwise MEM fails to fit the neuronal data accurately in our experiment and previous reports \cite{tkavcik2014searching, Ganmor2011}. This suggests that while microscopic dynamics of individual neurons are difficult to predict with sufficient accuracy at the whole-brain scale, their coarse-grained activity can be characterized by the interactions between neuronal populations or brain areas.  

In addition to modeling the probabilities of individual states, we also wanted to test whether the MEM can further characterize dynamics such as the transition between brain states. For this, we simulated brain activity sequences as a random walk on the energy landscape between neighboring brain states, as defined by the pairwise MEM. This simulation predicts the dwell time in individual attractor basins, as well as the frequency and probability of transitions between basins, with high accuracy. This finding suggests that resting state macroscopic brain state transitions are a natural consequence of the energy landscape over the states induced by the maximum entropy principle. 

The optimum scale of coarse-graining that allows for both accurate predictions of brain state probabilities and for the models to map onto known structural connectivity patterns likely depends on the exact nature of the data. MEM approaches have demonstrated excellent fits to dynamics ranging from spiking behavior in neuronal populations \cite{tang2008maximum} to whole-brain functional MRI patterns in the human brain \cite{watanabe2013pairwise}. The applicability of MEM approaches across such diverse scales suggests that pairwise interactions underlie dynamics at multiple scales, and that the optimum scale for a given dataset can be identified empirically through clustering or coarse-graining. 

There is a tension here in our understanding of brain dynamics: simple models can comprehensively describe large-scale brain activity patterns, yet neuronal assemblies encode temporally and spatially precise information in ways that can be modulated through the activity of only a few isolated neurons \cite{Carrillo-Reid2019-holographic}. Identifying how large-scale patterns of whole-brain dynamics, as described here, shape information processing and neuronal responses at the microscale is a current research area of much interest, and simple organisms whose neuronal activity can be mapped across scales play an important role in further exploring this relationship \cite{Kato2015-global, Marques2020-states}.  

\subsection{Macroscale pairwise interactions reveal underlying structural connectivity}

Both neuronal activity and pairwise synaptic connectivity can be measured in the larval zebrafish brain at single-cell resolution, which allows unprecedented access to functional and structural connectivity at brain-wide scales \citep{chen2018brain, Kunst2018-structural}. The analysis presented here adds to existing evidence indicating a strong correlation between structural measures of the strength of physical synaptic interactions and functional (i.e., model-derived) estimates of effective interactions. This structure-function coupling has been reported across diverse neuronal data recording modalities \cite{watanabe2013pairwise, ashourvan2017energy,ashourvan2021pairwise}, including calcium imaging in the larval zebrafish \cite{van2023neural}. In our analysis, pairwise interactions between clusters of neurons quantified in the $J_{ij}$ matrix closely resemble the strength of pairwise interactions between regions as measured by axon count (i.e., the structural connectivity). This effect is robust when modeling neuronal clusters spanning orders of magnitude in size. 

However, the model-based estimates of pairwise interactions do not capture underlying structural connectivity equally well across all parameter choices. Indeed, there appears to be a ‘characteristic scale’, in terms of number of neural clusters, at which the effective MEM interactions $J_{ij}$ performed best at identifying above-threshold structural connections between brain regions: At the relatively coarse scale of 12 clusters consisting of an average of $1623 (\pm 392)$ to $17740 (\pm 3122)$ neurons each, the majority of large between-region structural connections could be identified from the MEM interactions $J_{ij}$. There is, therefore, an apparently closer structure-function coupling at coarser scales of the analysis than at the neuronal level. This might indicate that averaging over some microscale heterogeneity at the neuronal level results in a mean-field approximation of the structural connectivity, which constrains the coarse-grained activity patterns accordingly.

Biologically, the majority of neuronal interactions are mediated through direct neuron-to-neuron directed signaling at the synapse \cite{Katz1999-neuralcomms}. However, other types of coupling -- including ephaptic transmission, extracellular fields, metabolic regulation \cite{Dudek1998-nonsynaptic, Ikeda2022-nonsynaptic} andmore complex patterns of synaptic connections between three or more neurons -- may deviate from a simple pairwise model. Many of the neuronal interactions that are not mediated through typical pairwise synaptic coupling are highly local, affecting mostly neighboring neurons rather than mediating fast, between-region interactions \cite{Su2012-nonsynaptic}. Thus, coarse-graining single-cell data to functional clusters may separate pairwise synaptic coupling that governs between-region connectivity from higher--order interactions that are subsumed in the within-cluster average. Furthermore, whilst single neuron dynamics are highly non-linear \cite{Herz2006-singleneuron}, coarse-grained brain dynamics are well-approximated by linear models \cite{Nozari2020-linear} – suggesting that coarse-graining may improve model performance for models based on linearity assumptions. 

\subsection{States of global hypersynchrony can emerge without coupling changes}

The healthy vertebrate brain spontaneously undergoes significant changes in whole-brain activity states characterized by varying levels of synchrony, for example, during the transition from wakefulness to slow-wave sleep \cite{Leung2019-sleep, Guo2022-sleep}. Similarly, in conditions like epilepsy, there are rapid paroxysmal shifts between more synchronous and less synchronous activation states. These rapid physiological and pathological transitions between asynchronous and hypersynchronous brain states occur without significant structural changes in the synaptic coupling.

Using the MEM model, we can show that modulation of intrinsic excitability (modeled here as a temperature parameter $T$), without changes in between-node coupling (i.e., without changes in the $J_{ij}$ matrix), can lead to a sudden transition between asynchronous and hypersynchronous states. Specifically, increasing the temperature in the model flattens the probability distribution, effectively reducing the influence of pairwise structure that favors low energy states and increasing the randomness in the activity patterns. These findings build on similar modeling of sleep-wake transitions in human brain data \cite{Nghiem2018-sws}. Biological systems may demonstrate similar transitions through diffusely distributed modulatory neurotransmission (e.g., through neuromodulators such as acetylcholine and noradrenaline that play a significant role in sleep/wake transitions).

Notably, the simulated activity patterns in the MEM-derived model of zebrafish brain's spontaneous activity do not show a gradual transition from asynchronous to hypersynchronous activity but rather show a sudden phase transition with the gradual change in the $T$ parameter resulting in significant changes of whole-brain activity patterns around the transition region. In systems poised at the boundary between these regimes, one would therefore expect the system behavior to be very sensitive to even small changes in the $T$ parameter. In fact, these findings confirm evidence of state transition dynamics during drug-induced seizures in larval zebrafish, which demonstrate that over short time scales, the dynamics can transition from critical to a chaotic, supercritical regime \cite{burrows2023microscale}.

\subsection{Alterations affect the response to perturbation}

The study of phase transitions and critical phenomena in regular lattices has led to key insights in statistical physics \citep{dorogovtsev2002ising,dorogovtsev2008critical}. However, biological networks like the brain have connectivity patterns that are unlike lattices. These complex networks  exhibit non-uniform connectivity patterns, such as power-law degree distributions in scale-free networks \cite{barabasi2009scale}. As a result, phase transitions and critical phenomena in complex networks are significantly different from those in regular lattices. The Ising model, in particular, exhibits anomalous phase transitions in complex scale-free \cite{dorogovtsev2002ising,aleksiejuk2002ferromagnetic,bianconi2002mean,leone2002ferromagnetic,bradde2010critical}, modular \cite{dasgupta2009phase,chen2011optimal,huang2015phase}, and core-periphery networks \cite{chen2018double}. 

For instance, unlike the first-order phase transitions of lattice structures, random scale-free networks display a second-order phase transition. In addition, the critical temperature for the onset of ferromagnetic ordering depends on the network's degree distribution, which determines the universality class of the phase transition \cite{dorogovtsev2002ising}. Moreover, modular structures in complex networks can give rise to metastable phases in both the equilibrium and nonequilibrium regimes, marked by the coherent alignment of intra-community spins and misaligned inter-community spins \cite{dasgupta2009phase,chen2011optimal,huang2015phase}. 

Consequently, modifying brain networks can have non-trivial effects on overall dynamics. For example, removing a single region from the zebrafish brain network can have drastic effects on the global dynamics. There is some previous evidence that inferences on global network dynamic outcomes can be drawn on \emph{virtual} resection simulations. For instance, densely connected regions tend to have an outsized influence and their removal results in significant changes in large-scale dynamics of the system as a whole \cite{khambhati2016virtual, Ren2019-nodedismantling, Moutsinas2020-noderemoval}. Moreover, resection analyses are starting to play a role in cognitive models, where they allow hypothesis generation for processes such as language, attention and memory \cite{honey2009predicting}. 

Similarly, resection analysis can also be used to study the robustness and vulnerability of networks to perturbations \cite{albert2000error}.Here, we examined the changes in the phase transition of the system toward globally hypersynchronous states, as a proxy for evaluating the resiliance of an altered system to pathological dynamics as seen for example during epileptic seizures. Specifically, evaluating post \emph{virtual resection} networks for changes in sensitivity to perturbation through specific heat and susceptibility meausres reveals node-specific changes to whole-network sensitivity to small perturbations following removal of single network nodes. In our model and previous finding these node-specific effects seem to be mediated through the nodes' particular connectivity patterns \cite{buldyrev2010catastrophic, gutierrez2011node}

These findings highlight the importance of functional hub brain regions in maintaining the overall function and resilience of the zebrafish brain network. The approach taken here allows us to demonstrate that the virtual resection of functional clusters involving the optic tectum has the largest impact on overall resultant network dynamics, mirroring its essential role in guiding zebrafish larval behaviour Previous studies have identified overlapping regions as central hubs maintaining resting state dynamics, with rich effective connectivity to most other brain regions \cite{rosch2018calcium}. Our framework now provides a modeling approach to infer network structure and behavior from dynamic observations of complex systems. Similar approaches in future may inform the development of targeted interventions in complex networks, such as surgery or regional neuromodulations for neurological and psychiatric disorders, such as epilepsy \cite{khambhati2016virtual, Wang2023-virtualbrain, Lehnertz2023-prediction}, or disorders of consciousness \cite{hudetz2014spin,kandeepan2020modeling,abeyasinghe2020consciousness}. In addition, our study demonstrates the usefulness of resection analysis in studying the functional connectivity of complex networks, capable of providing valuable and empirically verifiable predictions. 

\appendix

\section{Datasets}

In order to capture whole-brain activity at cellular resolution in a vertebrate brain, we used resting-state whole-brain calcium imaging datasets recorded in larval zebrafish at day 6 post-fertilization using light sheet imaging. Dataset acquisition is explained in detail in \cite{Chen2018-zebra} and was accessed through the public repository \cite{Chen2019-dataset}. Briefly, images were recorded using a light sheet microscope at 2 volumes/second in transgenic larval zebrafish expressing the genetically encoded calcium indicator \textit{GCAMP6f} pan-neuronally within the cell nucleus. This approach allowed automatic cell detection \cite{Kawashima2016-zebra}, resulting in approximately ~80,000 individual neuron calcium fluorescence traces per fish. For the analysis here we used resting-state segments, recording spontaneous activity under homogeneous background illumination conditions without changes in visual stimuli for approximately 5-8 minutes recorded for 11 individual larvae. Individual zebrafish brains were registered to an atlas template \cite{Randlett2015-atlas}, and standardized atlas-based locations are used for subsequent analyses. 

\section{Pairwise maximum entropy model}
\addcontentsline{toc}{subsection}{Pairwise maximum entropy model}

To bridge the gap between microscopic details and macroscopic predictions about the likelihood of different neural states, we employ a pairwise maximum entropy model (MEM). By maximizing the entropy of the distribution over activity states, we arrive at a prediction that is optimally unbiased, given a set of microscopic details about the system. Here, we constrain the average activities and pairwise correlations between different brain regions, thus yielding the pairwise MEM. To fit the pairwise MEM, we calibrate the external fields $h_i$ and functional interactions $J_{ij}$ such that the model matches the activation and co-activation rates observed in the real system. From these microscopic constraints, the pairwise MEM makes large-scale predictions about the large-scale zebrafish brain dynamics.

We applied dimensionality reduction to all fish datasets using k-means clustering based on cell position. This process resulted in the identification of 1000 clusters of neurons, determined by their spatial proximity. Subsequently, these spatial clusters were grouped into a smaller number of functional clusters (ranging from 8-16, 32, 64, and 128) by analyzing them based on their covariance. The covariance between clusters was calculated from the average calcium traces of all neurons within each cluster.

We analyzed 83 minutes of recordings from 11 zebrafish larvae. At each time point $t$, the activation state of the system is defined by the binary vector $\sigma^{t}=\left [ \sigma_1^t , \sigma_2^t ,..., \sigma_N^t \right ]$, where $\sigma_{i}^t$ is the binarized average calcium trace of cluster $i$ at time $t$ and $N$ is the total number of clusters. Specifically, $\sigma_{i}^t=1$ (-1) for activation above (below) the z-scored average calcium time series at various thresholds (Z = 0, 0.5, 1, 1.5, 2). To obtain the empirical activation rate of cluster $i$, we calculated the average of $\sigma_i^t$ over all time slices. This is represented as $\left \langle \sigma_i \right \rangle = \frac{1}{T} \sum_{t=1}^\tau \sigma_i^t$, where $\tau$ is the number of time slices. Similarly, the empirical correlation between cluster $i$ and $j$, denoted as $\left \langle \sigma_i \sigma_j \right \rangle$, is defined as the average of the product of $\sigma_i^t$ and $\sigma_j^t$ over all time slices, which is calculated as $\textstyle \frac{1}{\tau} \sum_{t=1}^\tau \sigma_i^t \sigma_j^t$.

The pairwise MEM is constrained such that the model averages $\left \langle \sigma_i \right \rangle_{m}$ and $\left \langle \sigma_i\sigma_j \right \rangle_{m}$ match the empirical values of $\left \langle \sigma_i \right \rangle$ and $\left \langle \sigma_i \sigma_j \right \rangle$. The probability distribution over states that satisfies these constraints and maximizes the entropy is the Boltzmann distribution \citep{jaynes1957information}:
\begin{equation}
P(\sigma)=\frac{1}{Z}e^{-E(\sigma)},
\label{dyn1}
\end{equation}
where $Z$ is the normalizing partition function, given by:
\begin{equation}
Z=\sum_{\sigma} e^{-E(\sigma)},
\end{equation}
and $E(\sigma)$ is the energy of this state, given by:
\begin{equation}
E(\sigma) = - \sum_{i=1}^{N} h_{i} \sigma_{i} - \frac{1}{2} \sum_{i,j=1}^{N} J_{ij}\sigma_{i}\sigma_{j}.
\label{dyn2}
\end{equation}
The parameters $h_{i}$ represent the bias towards activation and $J_{ij}$ represent the functional interaction between clusters $\textit{i}$ and $\textit{j}$. To fit the pairwise MEM, one adjusts the parameters $h_{i}$ and $J_{ij}$ using a gradient descent algorithm \citep{watanabe2014energy} until the empirical averages $\left \langle \sigma_i \right \rangle$ and $\left \langle \sigma_i \sigma_j \right \rangle$ match the model averages.

%Since the state space is prohibitively large in our data, we followed \citep{gu2018energy} by approximating the model averages $\left \langle \sigma_i  \right \rangle_{m}$ and $\left \langle \sigma_i\sigma_j  \right \rangle_{m}$ by calculating the correlation of a sequence of random samples ($N=10,300,000$, after discarding the first 300,000 and downsampling by a factor of 500) from the probability distribution using the Metropolis--Hastings algorithm \citep{metropolis1953equation}. 
\

The aforementioned methods are computationally expensive when the dimensionality is higher than $N=15$. To overcome this hurdle, for higher dimensions, we used a pseudo-likelihood maximization algorithm instead of the likelihood maximization approach. The MATLAB scripts provided by \cite{ezaki2017energy} were used to estimate the model parameters for dimensions larger than $N=15$. In the pseudo-likelihood maximization scheme, the goal is to solve the following equation:
\begin{equation}
(h,J)=\operatorname*{argmax}_{h,J}\mathcal{L}(h,J),
\label{dyn3}
\end{equation}
\noindent where the pseudo-likelihood function, $\mathcal{L}(h,J)$, is defined as
\begin{equation}
\mathcal{L}(h,J) \approx \prod_{t=1}^{t_{max}}\prod_{i=1}^{N}\tilde{P}(\sigma_{i}(t)|h,J,\sigma_{/i}(t))),
\label{dyn4}
\end{equation}
\noindent where $\tilde{P}$ represents the Boltzmann distribution for a single spin (i.e., cluster) $\sigma_{i}$ given that the other $\sigma_{j}$($j$$\neq$$i$) values are fixed to $\sigma_{/i}(t)$ $\equiv $ $(\sigma_{1}(t), ... , \sigma_{i-1}(t), \sigma_{i+1}(t), ... , \sigma_{N}(t))$. Therefore, $\tilde{P}$ is given by
\begin{equation}
\tilde{P}(\sigma_{i}(t)|h,J,\sigma_{/i}(t))) = \frac{e^{(h_i\sigma_i+\sum_{j=1}^{N} J_{ij}\sigma_{i}\sigma_j(t)))}}{\sum_{\sigma'_{i} = 1,0} e^{(h_i\sigma'_i+\sum_{j=1}^{N} J_{ij}\sigma'_{i}\sigma_j(t)))}} ,
\label{dyn5}
\end{equation}

The above method uses a mean-field approximation that disregards the influence of one variable on another. However, the estimator obtained by maximizing the pseudo-likelihood converges to the maximum-likelihood estimator as the number of time steps $t_{max}$ increases, as per \cite{besag1975statistical}. To estimate the model's parameters, $h$ and $J$, we use a gradient descent scheme, which updates the parameters by comparing the empirical mean and correlation values to the mean and correlation values predicted by the model. The update equations are as follows:
\begin{equation}
h_{i}^{new} - h_{i}^{old} = \epsilon (\left \langle \sigma_i \right \rangle _{empirical} - \left \langle \sigma_i \right \rangle _{\tilde{P}} )
\label{dyn6}
\end{equation}
and
\begin{equation}
J_{ij}^{new} - J_{ij}^{old} = \epsilon (\left \langle \sigma_i \sigma_j \right \rangle _{empirical} - \left \langle \sigma_i \sigma_j \right \rangle _{\tilde{P}}),
\label{dy7}
\end{equation}
\noindent where the superscripts new and old represent the parameters after and before a single updating step, respectively, $\epsilon >0$ is the learning rate, and $\left \langle \sigma_i \right \rangle _{\tilde{P}}$ and $\left \langle \sigma_i \sigma_j \right \rangle _{\tilde{P}}$ are the mean and correlation with respect to distribution $\tilde{P}$ (equation (5)) and are given by
\begin{equation}
 \left \langle \sigma_i \right \rangle _{\tilde{P}} =\frac{1}{t_{max}} \sum_{t=1}^{t_{max}} \tanh\left [ h_i+\sum_{j'=1 }^{N} J_{ij'} \sigma_{j'}(t)  \right ]
\label{dyn8}
\end{equation}
and
\begin{equation}
 \left \langle \sigma_i \sigma_j \right \rangle _{\tilde{P}} =\frac{1}{t_{max}} \sum_{t=1}^{t_{max}} \sigma_j(t) \tanh\left [ h_i+\sum_{j'=1 }^{N} J_{ij'} \sigma_{j'}(t)  \right],
 \label{dyn9}
\end{equation}
respectively. For more details regarding the likelihood and pseudo-likelihood maximization algorithms and scripts, see \cite{ezaki2017energy}.

% We can also represent the activation states of neuronal clusters using $\tilde{\sigma_{i}} \in \left \{ -1 ,1 \right \}, (i=1,...,N)$, which is mathematically equivalent to using $\sigma_{i} \in \left \{ 0 ,1 \right \}$. This means that the relationship between the two is $ 2\sigma_{i}-1=\tilde{\sigma_{i}}$. As a result, the pairwise MEM parameters, $J_{ij}$ and $h_{i}$, also have the following relationships:
% \begin{equation}
% h_{i}= 2\tilde{h_{i}} -2\sum_{i=1}^{N}\tilde{J_{ij}}
% \label{dyn10}
% \end{equation}
% \begin{equation}
% J_{ij}= 4\tilde{J_{ij}}
% \label{dyn11}
% \end{equation}
% We used the method provided by \cite{ezaki2017energy} to estimate the model parameters with the assumption that the states are represented by $\tilde{\sigma_{i}} \in \left \{ -1,1 \right \}$. To align our results with this assumption, we used equations \ref{dyn10} and \ref{dyn11} to transform the estimated $\tilde{J_{ij}}$ and $\tilde{h_{i}}$ matrices.

%Usually, calculating the normalization constant in the Boltzmann distribution (i.e., \emph{partition function}) in high-dimensional datasets is very difficult, as it involves summation over all possible states. This transformation allows us to approximate the partition function ($\hat Z$) directly from the probability of the \emph{silent} state (i.e., $p(000\cdots0)$) as $\hat Z=\frac{1}{p(000\cdots0)}$, since it holds that $ p(000\cdots0)=\frac{1}{Z}$  \cite{Ganmor2011,tkavcik2014searching}.

\section{Dimensionality reduction and structure-function coupling}
\addcontentsline{toc}{subsection}{Dimensionality reduction and structure-function coupling}

The correlation between structural and functional connectivity is a defining characteristic of brain networks in  humans \cite{sporns2005human,honey2009predicting} and other species (e.g., mice \cite{Wang2015}, rats \cite{Smit2013}, and monkeys \cite{Markov2014}). Recent studies have provided further evidence of structural and functional similarity in the zebrafish brain \cite{van2023neural}. In light of these observations, we aimed to reduce the dimensionality of the functional dataset by selecting a resolution of functional parcellations that maximizes the similarity between the estimated $J_{ij}$ and the structural connectivity matrix (total tract count between large-scale clusters). Our findings reveal the high similarity between structural and functional matrices across 8-16, 32, 64, and 128 scales. However, the highest similarity, as measured by the correlation between the two matrix modalities and the AUC for the detection of binarized structural connections between regions from $J_{ij}$ weights, was consistently identified at $N= 12$ regions and functional binarization threshold of 1 and 30 percent of the maximum weight of the structural connectivity matrix (see SI Fig. 4).

\section{Evaluating the accuracy of the pairwise MEM}
\addcontentsline{toc}{subsection}{Evaluating the accuracy of the pairwise MEM}

We used information-theoretic approaches to evaluate the fit of our model. A pairwise maximum entropy model (MEM) better fits observed dynamics than a first-order (independent) model since it considers not only regional activations but also pairwise correlations between regional time series. These additional considerations lead to a lower uncertainty or entropy in the second-order model than in the first-order model. Intuitively, increasing the order of the model will monotonically decrease the entropy closer to the true entropy, which can be measured empirically. We can measure this entropy difference using the multi-information, $I_N=S_1-S_N$, given by the difference between the first-order model entropy, $S_1$, and the empirical entropy of the data, $S_N$. In the context of our study, the multi-information measures the total amount of correlation in brain signals independent of higher-order correlations. To evaluate the performance of the pairwise MEM, we asked whether the reduction in entropy following the incorporation of pairwise interactions in the model, $I_2=S_1-S_2$, captured the majority of the total multi-information. In other words, we quantified the performance of the pairwise MEM as the fraction of the multi-information captured by the second-order model, $r =I_2/I_N$, where $r$ can range between 0 and 1. Our results showed that pairwise interactions accounted for a large portion of the multi-information, approximately $84 \%$. 

\section{Identifying local minima and basins of the energy landscape}
\addcontentsline{toc}{subsection}{Identifying local minima and basins of the energy landscape}

We defined the energy landscape by the network of cluster activation states and their energy values. In this landscape, the neighboring states are only one hamming distance apart, meaning that adjacent states are identical up to the activity of one brain region \citep{fninf.2014.00012}. To identify the local minima or attractor states, using a steep search algorithm we exhaustively searched the entire landscape to find the states with energies lower than all their adjacent states. Next, to find the states that belong to the basin of each local minima, we first start at a given state $\sigma$ and iteratively move to a neighboring state $\sigma'$ if $E(\sigma') < E(\sigma)$. We continue tracing this path until we reach a local minimum where no neighboring states have lower energy (similar to \citep{watanabe2014energy,watanabe2013pairwise}). We consider the basin size of this local minimum to be the ratio between the number of basin states to the total number of possible states.

We were particularly interested in identifying the obstacles that impede the transition between the different basins of attraction. To investigate these transitions, we employed a method that involved removing the highest energy state from the energy landscape, along with the edges linking it to its neighboring states. We then determined whether each pair of local minima were still connected by a path in the reduced landscape. We repeated this process until we discovered the state whose removal causes one or more local minima to be disconnected within the landscape (i.e., the saddle states). We repeated this process until we arrived at a reduced landscape where all the local minima were disconnected, and we were able to identify all the saddle states. The disconnectivity graph in Fig. \ref{figure_4}a was created using the identified energy values of the identified local minima and saddle states. 

We calculated the \textit{asymmetric} energy barrier \citep{zhou2011random} between each pair of local minima by taking the difference between the energy of the saddle state and the energies of the two minima. We then defined the \textit{symmetric} energy barrier between two local minima as the minimum between the two asymmetric energy barriers. If the energy barrier between two local minima is high, then we hypothesize that the rate of transition between them is low, at least in one direction \citep{watanabe2013pairwise}. However, our results did not show a strong relationship between estimated energy barriers (symmetric and asymmetric) between local minima states and the empirical basin transition probabilities (SI Fig. 8). 

\section{Simulating state transitions}
\addcontentsline{toc}{subsection}{Simulating state transitions}

To better understand the spontaneous patterns of activations and state transitions, we simulated the large-scale dynamics as a random walk process over the estimated energy landscape using a Markov chain Monte Carlo with Metropolis-Hastings (MCMC) algorithm \citep{metropolis1953equation,hastings1970monte,zhou2011random}. In this method, the activation state $\sigma$ is allowed an isometric transition to one of $N$ neighboring states. Next the actual transition from $\sigma$ to $\sigma'$ occurs with probability $\textstyle P(\sigma' | \sigma) = \min\left [ 1,e^{E(\sigma)-E(\sigma')} \right ]$. We conducted a $10^{6}$ step (plus $3\time10^{4}$ initial steps) walk with randomly chosen initial states. Next, we removed the initial steps to thermalize and ensure the independence of results from the initial conditions. Finally, we construct a trajectory between different basins of attraction from state walks. By comparing the basin transition probabilities in simulations and experiments, we can assess the extent to which the dynamics of the proposed random walk model align with those observed in the brain.

\section{Evaluating the thermal properties of the system}
\addcontentsline{toc}{subsection}{Evaluating the thermal properties of the system}

The thermal behavior of the MEM was simulated using a MCMC algorithm. We account for the temperature of the system using the following variant of the Boltzmann distribution over activity states \ref{dyn2}:
\begin{equation}
P(\sigma)= \frac{1}{Z} e^{-\frac{1}{T} E(\sigma)}
\label{dyn1T}
\end{equation}
where $T$ is the temperature and the partition function is now given by:
\begin{equation}
Z=\sum_{\sigma} e^{-\frac{1}{T}E(\sigma)}.
\end{equation} We initialized the simulation with a random initial state followed by 1030000 random walks. Similar to the above-mentioned state transition simulations, The first step of the simulation involved discarding the initial 30000 steps as thermalization steps and then down-sampling the remaining walks every 500 samples. The system's final state at each temperature step was then used as the initial state for the next lower temperature step. The simulations were repeated at different temperatures, starting at the highest temperature of $T= 2$ and gradually reducing the temperature to $T=0$. 

We used the fluctuation-dissipation theorem \cite{} to calculate the specific heat and susceptibility. This theorem relates the fluctuations of the energy to the changes in temperature, such that the specific heat is given by:
\begin{equation}
C=\frac{\left \langle E^{2} \right \rangle - \left \langle E \right \rangle ^{2} }{T^{2}},
\label{SH}
\end{equation}
where $E$ is the energy, $T$ is the temperature, and angle brackets indicate an average in the MEM. Similarly, the susceptibility is given by:
\begin{equation}
\chi =\frac{\left \langle M^{2} \right \rangle - \left \langle M \right \rangle ^{2} }{T},
\label{SU}
\end{equation}
where $M = \sum_i \sigma_i$ is the magnetization of the system.

To understand the role of different regions in the system's thermal behavior we removed, one by one, each region its connections. We then compared the specific heat and susceptibility curves obtained from the resection simulation to the original system to understand how the resected region changes the critical behavior of the system. Specifically, we measured the shift in the peak of the specific heat and susceptibility curves. 

\bibliography{journalBiblio}
\bibliographystyle{naturemag} 

\end{document}

\preprint{APS/123-QED}

\title{Supplementary document for "Spontaneous brain activity emerges from pairwise interactions in the  larval zebrafish brain"}

\author{Richard E. Rosch}
\affiliation{Department of Clinical Neurophysiology, King's College Hospital NHS Foundation Trust, London, United Kingdom}
\author{Dominic R. W. Burrows}
\affiliation{MRC Centre for Neurodevelopmental Disorders, King's College London, London, United Kingdom}
\author{Christopher W. Lynn}
\affiliation{Initiative for the Theoretical Sciences, Graduate Center, City University of New York, New York, USA}
\affiliation{Joseph Henry Laboratories of Physics and Lewis–Sigler Institute for Integrative Genomics, Princeton University, Princeton, NJ, USA}
\author{Arian Ashourvan}
\affiliation{Department of Psychology, University of Kansas, Lawrence, United States of America}
\email{Correspondence to: ashourvan@ku.edu}

\date{March 2023}

\maketitle

\tableofcontents

\newpage

\section*{SI1. Evaluation of k-means clustering across different cluster sizes.}
%\addcontentsline{toc}{section}{SI1. Joint-estimation algorithm}

%\section{\label{sec:level1}Introduction}

We used several criteria for evaluating the quality of k-means functional clusters. These criteria help to assess how well the data points are separated into distinct clusters and how meaningful these clusters are. Therefore, these clustering evaluation criteria can help determine the optimal number of clusters and assess the quality of the clustering results.

\subsection*{Calinski-Harabasz criterion}
We utilized the Calinski-Harabasz criterion \citep{Calinski1974}, which is also known as the variance ratio criterion, to determine the optimal number of clusters in our data. This criterion is calculated using the following formula:

\begin{equation}
VRC=\frac{SS_{B}}{SS_{W}} \times \frac{(N-k)}{(k-1)},
\end{equation}

where $k$ represents the number of clusters, $N$ is the total number of observations, and $SS_{B}$ and $SS_{W}$ are the between-cluster and within-cluster variances, respectively. The equations for $SS_{B}$ and $SS_{W}$ are:

\begin{equation}
SS_{B} =\sum_{i=1}^{k}n_{i} \left | m_{i}-m \right |^{2},
\end{equation}

\begin{equation}
SS_{W} =\sum_{i=1}^{k} \sum_{x\in c_{i}}^{} \left | x-m_{i} \right |^{2},
\end{equation}

where $n_{i}$ is the number of points in the $i_{th}$ cluster with the centroid $m_{i}$, $m$ is the total mean of the data, $x$ represents a data point, $c_{i}$ represents the $i_{th}$ cluster, and $\left | x-m_{i} \right |$ and $\left | x-m_{i} \right |$ are the Euclidean distances ($L^{2}$ norm) between the two vectors. The Calinski-Harabasz criterion aims to maximize the variance ratio criterion by identifying the optimal number of clusters with high between-cluster variance and low within-cluster variance.

\subsection*{Davies-Bouldin criterion}

We also used the Davies-Bouldin criterion \citep{Davies1979} to determine the optimal number of clusters. This criterion captures the ratio of within- and between-cluster distances and is calculated using the following formula:

\begin{equation}
DB =\frac{1}{k}\sum_{i=1}^{k}\max_{j\neq i}{D_{i,j}},
\end{equation}

where $D_{i,j}$ is the ratio of within-to-between cluster distance for clusters $i$ and $j$, and is defined as:

\begin{equation}
{D_{i,j}} =\frac{(\bar{d_{i}}-\bar{d_{j}})}{d_{i,j}},
\end{equation}

where $\bar{d_{i}}$ and $\bar{d_{j}}$ are the average distances between each data point in the $i_{th}$ and $j_{th}$ clusters to their own cluster centroids, and $d_{i,j}$ is the Euclidean distance between the centroids of clusters $i$ and $j$. The optimal number of clusters is identified by minimizing the Davies-Bouldin index, representing the best within-to-between cluster distance ratio.

\subsection*{Silhouette Analysis}

Silhouette Analysis, introduced by \cite{Rousseeuw1987}, is a method to evaluate the quality of clustering by measuring how similar each point is to other points within its cluster compared to points in other clusters. It is defined as follows:

\begin{equation}
S_{i}=\frac{a_{i}-b_{i}}{\max(a_{i},b_{i})},
\end{equation}

\noindent Here, $a_{i}$ represents the average distance between the $i_{th}$ data point and other points in its cluster, and $b_{i}$ represents the minimum average distance between the $i_{th}$ data point and points in different clusters (minimized across all clusters). A high Silhouette score, ranging between 1 and -1, indicates that the data point is well-clustered within its cluster and poorly matches the data points from other clusters. Conversely, many data points with zero or negative Silhouette values indicate the presence of few or many clusters in the data.

\section*{Supplementary Information (SI) figures}
%\addcontentsline{toc}{section}{SI Figures}

\begin{figure*}%[hbtp]
\centering
\includegraphics[width=0.8\linewidth ]{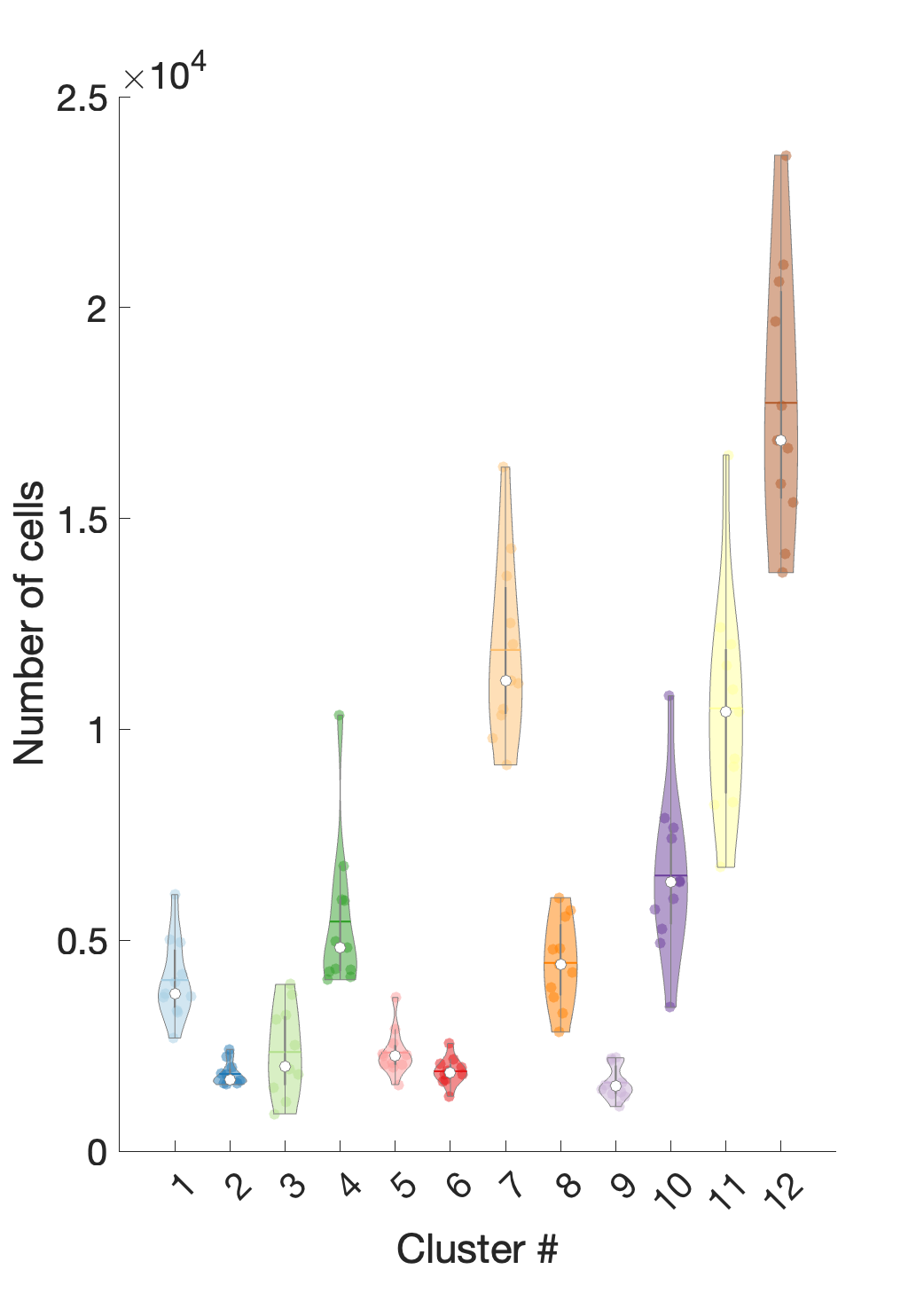}
\caption{\textbf{Distributions of the number of cells for all 12 functional clusters.} The violin plot displays the distribution of data points representing the number of cells for each zebrafish. The width of the violin at each point represents the density of data at that point, with the broadest part indicating the highest data density. Vertical lines and white dots within each violin represent the distribution's mean and median, respectively. } 
\label{SI_figure_1}
\end{figure*}

\begin{figure*}%[hbtp]
\centering
\includegraphics[width=1\linewidth ]{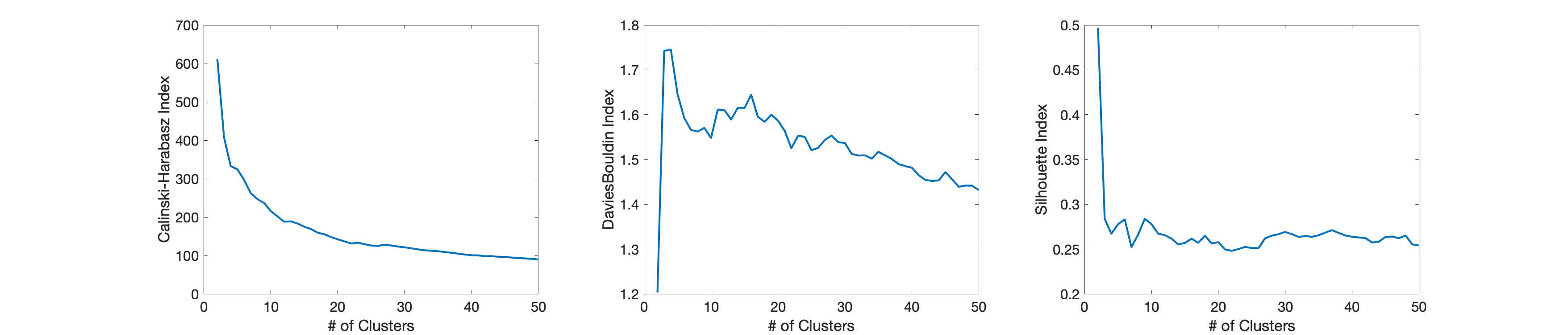}
\caption{\textbf{Evaluation of functional clusters across different sizes.} The  Calinski-Harabasz, Davies-Bouldin, and Silhouette indexes for different $k-$means cluster sizes from left to right plots, respectively. The results show that all three methods identify $N=2$ as the optimal cluster size with no clear converging results for $N>2$ clusters across the three methods. See the SI1 section for details about the aforementioned cluster evaluation criteria. } 
\label{SI_figure_2}
\end{figure*}

\begin{figure*}%[hbtp]
\centering
\includegraphics[width=1\linewidth ]{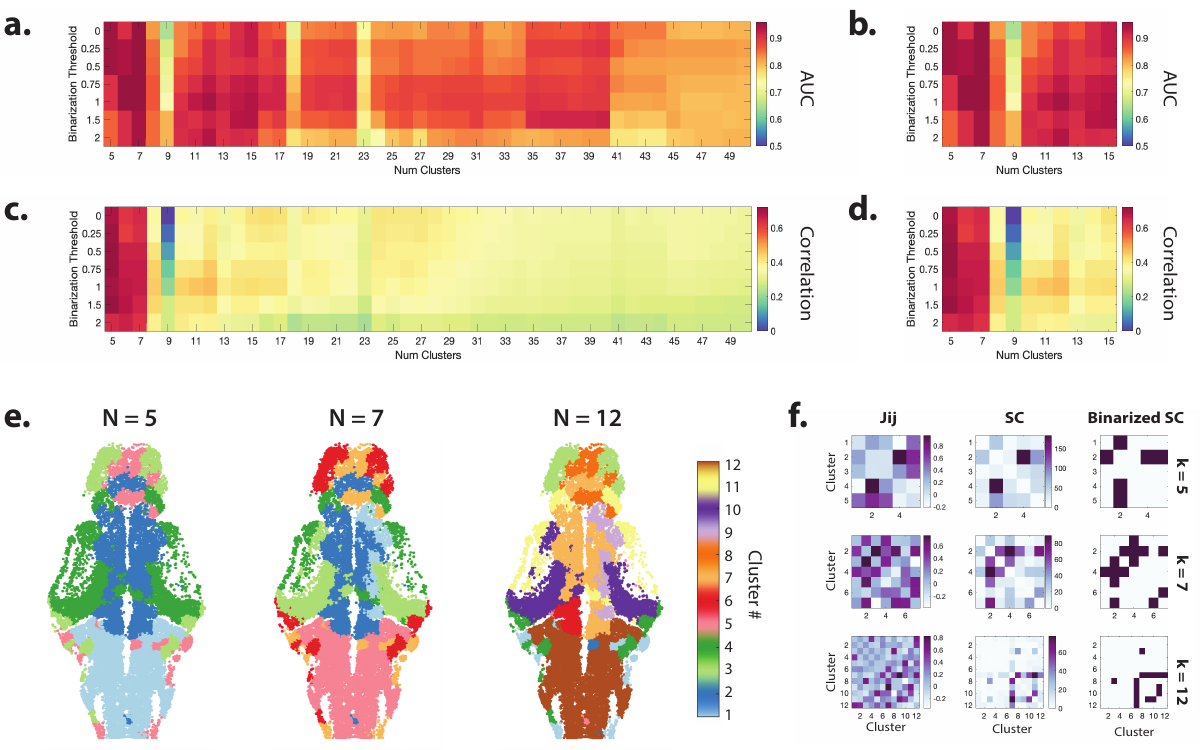}
\caption{\textbf{Similarity between the Structural and functional connectivity matrices across topological scales.} \textbf{a.} AUC values for detection of thresholded structural edges ($>30\%$ of maximum fiber count) based on $J_{ij}$ matrix estimated using pseudo-likelihood maximization algorithm based on clusters of different sizes and at different binarization thresholds of average cluster activations. \textbf{b.} Same as \textbf{a}, except the $J_{ij}$ matrix estimated using likelihood maximization algorithm. \textbf{c.} The correlation values between the structural edge weights and the $J_{ij}$ weights estimated using the pseudo-likelihood maximization algorithm based on clusters of different sizes and at different binarization thresholds of average cluster activation. \textbf{d.} Same as \textbf{c}, except the $J_{ij}$ matrix estimated using likelihood maximization algorithm. \textbf{e.} Detail of single z-slice demonstrating the spatial location of 5,7, and 12 functional clusters. Note that both $N=5$ and $N=7$ clusters identify several spatially disconnected clusters in the anterior-posterior loci. For example, clusters 3 and 5 for $N=5$ and 6 and 7 for $N=7$. \textbf{f.} The $J_{ij}$ matrices estimated using the likelihood maximization algorithm for high binarization thresholds ($z=1.5$) for $N=5$,7, and 12 clusters. The middle column shows the corresponding structural connectivity (SC) matrices (i.e., fiber count between clusters), and the left column shows the SC matrices thresholded at $30\%$ of maximum SC values.} 
\label{SI_figure_3}
\end{figure*}

% \begin{figure*}%[hbtp]
% \centering
% \includegraphics[width=0.5\linewidth ]{SC_Jij_MaxLikelihood_Correlation_Num_cluster_to_Bin_thresh.png}
% \caption{\textbf{.}  } 
% \label{SI_figure_3}
% \end{figure*}

\begin{figure*}%[hbtp]
\centering
\includegraphics[width=1\linewidth ]{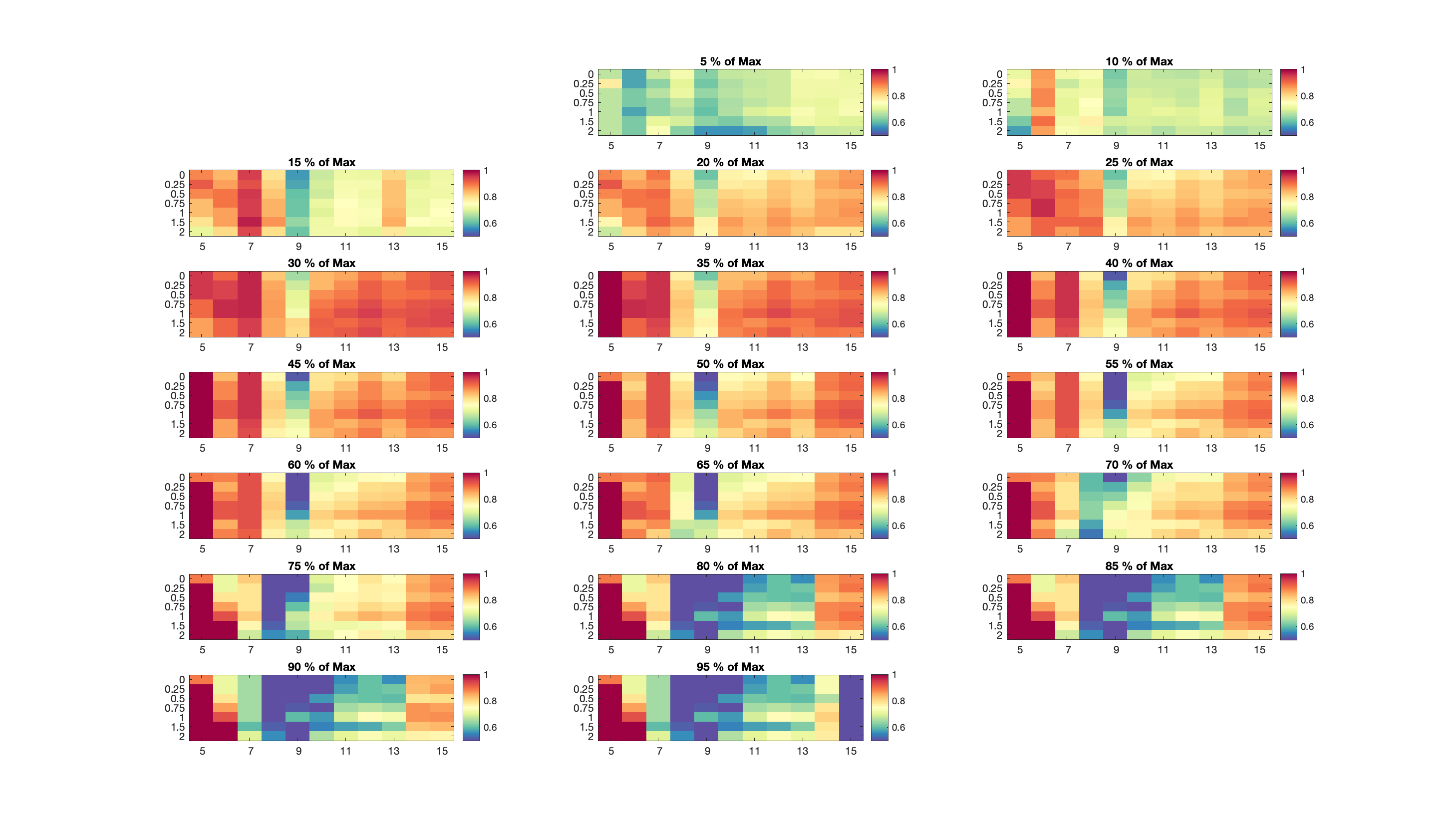}
\caption{\textbf{Accuracy of detecting structural connections using functional connectivity. } \textbf{a.} AUC values for detection of thresholded structural edges based on $J_{ij}$ matrix estimated using likelihood maximization algorithm based on clusters of different sizes ($N=5$ to 15) and at different binarization thresholds of average cluster activations. We provided the results for SC matrices thresholded at 5 to 95 $\%$ of maximum SC values.} 
\label{SI_figure_4}
\end{figure*}

\begin{figure*}%[hbtp]
\centering
\includegraphics[width=1\linewidth ]{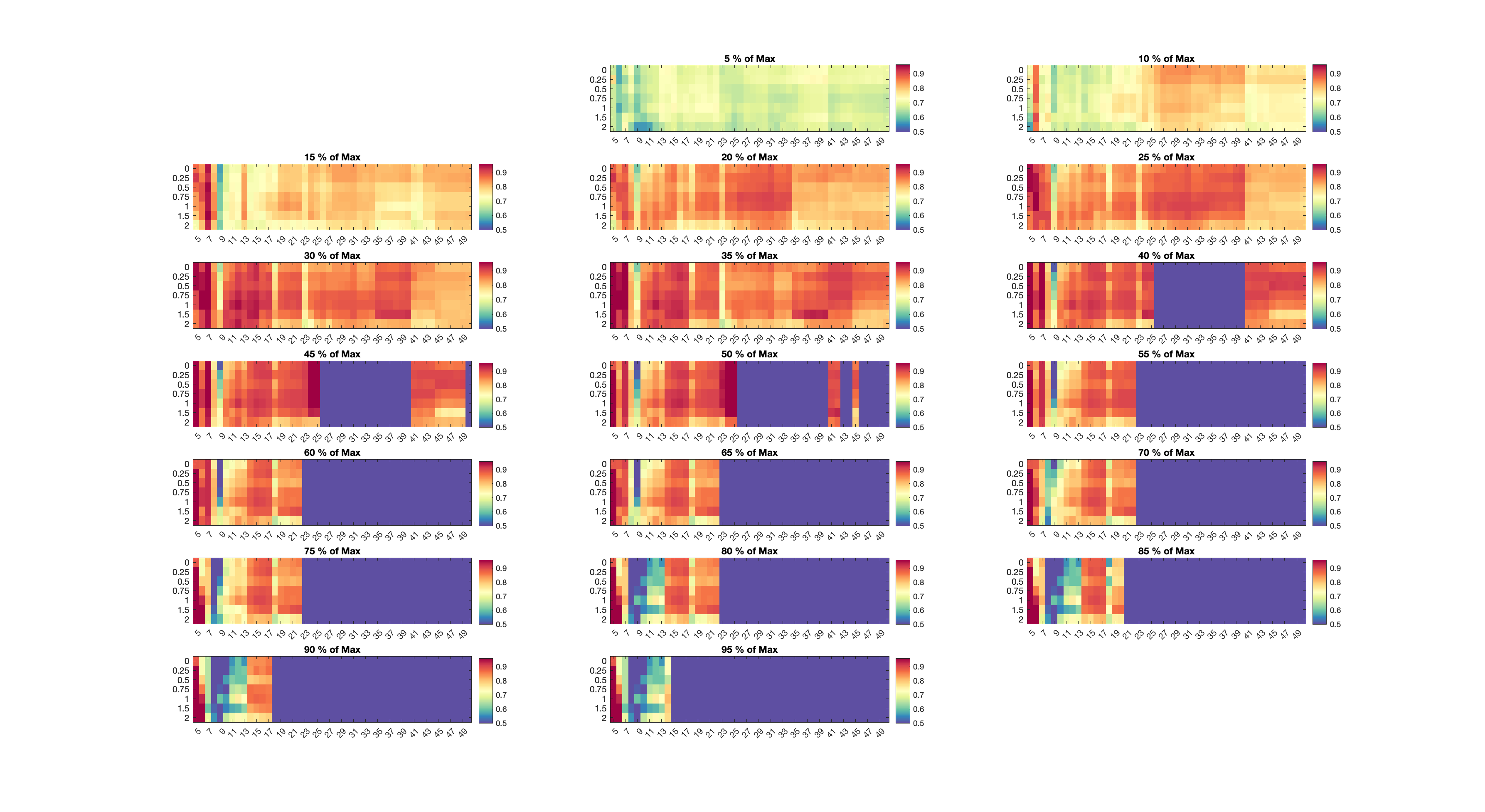}
\caption{\textbf{Accuracy of detecting structural connections using the functional connectivity estimated using pseudo-likelihood maximization scheme. } \textbf{a.} AUC values for detection of thresholded structural edges based on $J_{ij}$ matrix estimated using pseudo-likelihood maximization algorithm based on clusters of different sizes ($N=5$ to 15) and at different binarization thresholds of average cluster activations. We provided the results for SC matrices thresholded at 5 to 95 $\%$ of maximum SC values.} 
\label{SI_figure_5}
\end{figure*}

% \begin{figure*}%[hbtp]
% \centering
% \includegraphics[width=0.5\linewidth ]{SC_Jij_AUC_Num_cluster_to_Bin_thresh_Sweep_Max.png}
% \caption{\textbf{.}  } 
% \label{SI_figure_5}
% \end{figure*}

% \begin{figure*}%[hbtp]
% \centering
% \includegraphics[width=0.5\linewidth ]{SC_Jij_Correlation_Num_cluster_to_Bin_thresh.png}
% \caption{\textbf{.}  } 
% \label{SI_figure_6}
% \end{figure*}

\begin{figure*}%[hbtp]
\centering
\includegraphics[width=0.7\linewidth ]{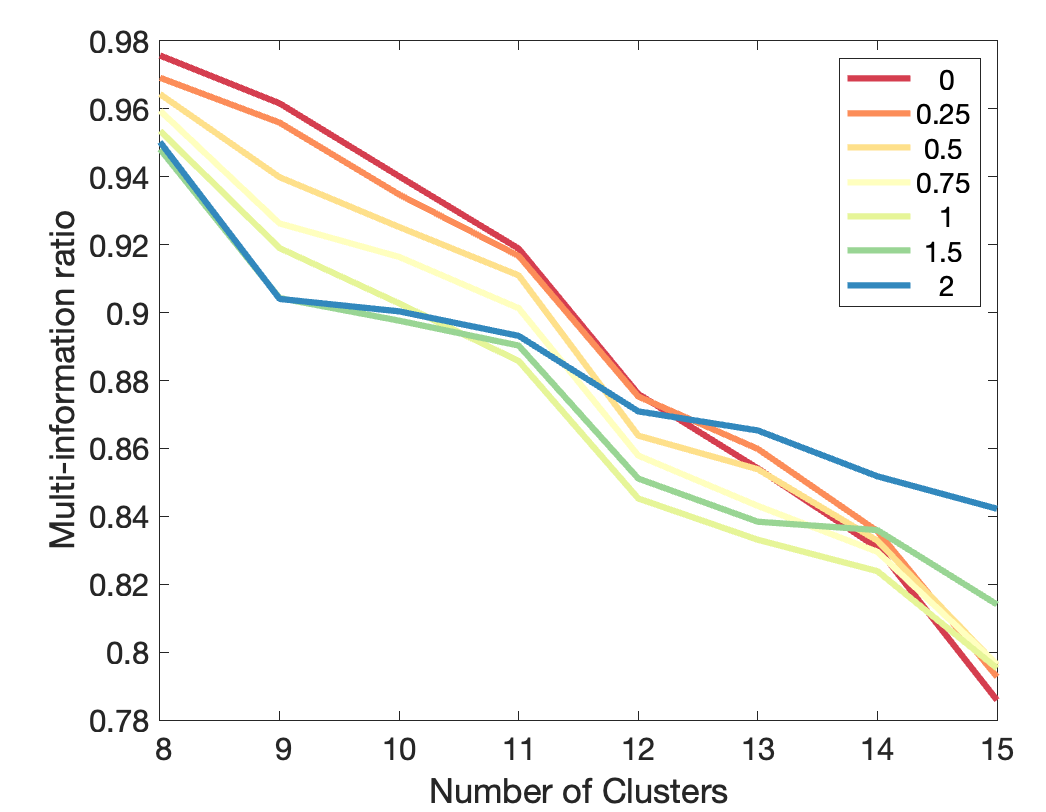}
\caption{\textbf{Goodness-of-fit of pairwise MEM and the number of clusters.} Here, we show the goodness-of-fit of the pairwise MEM  across different cluster sizes and the average cluster activation binarization threshold (color-coded) using the Multi-information ratio. See materials and methods for details on these metrics.} 
\label{SI_figure_6}
\end{figure*}

\begin{figure*}%[hbtp]
\centering
\includegraphics[width=1\linewidth ]{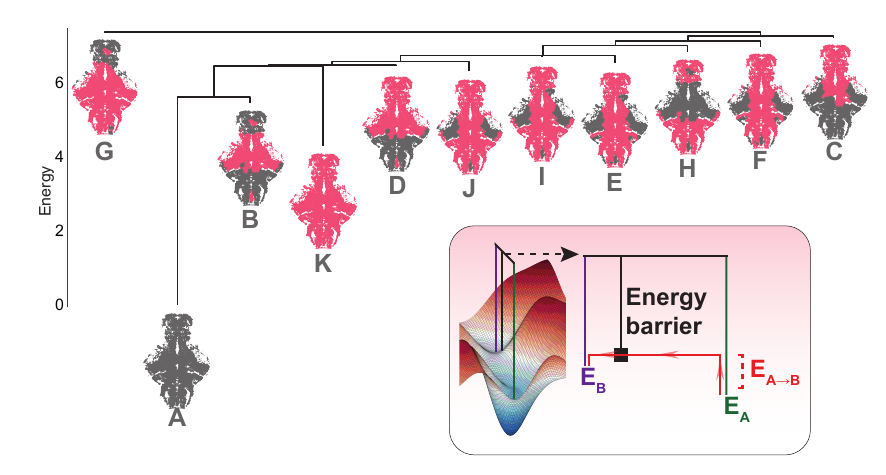}
\caption{\textbf{Disconnectivity graph between the local minima of the energy landscape.} \textbf{a.} Disconnectivity graph between the local minima of the energy landscape as defined in Fig. \ref{figure_1}b. We also illustrate the energy barrier between two example minima (1 and 7), defined by the difference in energy with respect to the saddle point state connecting the two minima.}
\label{SI_figure_7}
\end{figure*}

\begin{figure*}%[hbtp]
\centering
\includegraphics[width=1\linewidth ]{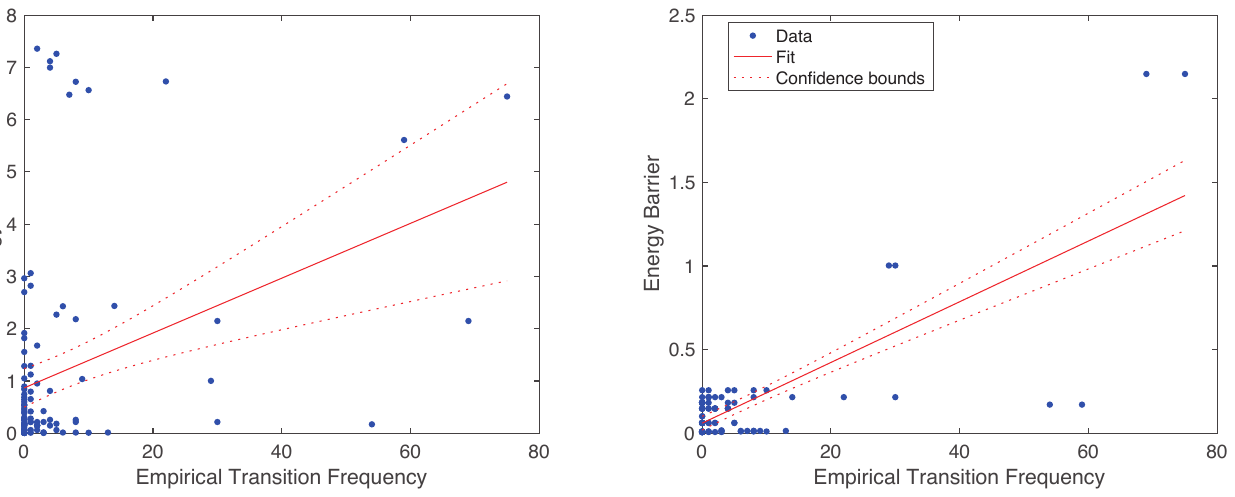}
\caption{\textbf{ The energy barrier between local minima and the transition frequency between them.} (Right) The energy barrier size and the frequency of transitions between the basins of the 11 local minima identified for $N = 12$ clusters.  (Left) The same as the right panel except that the y-axis represents the average (i.e., symmetric) energy barrier for each state transition. The solid and dashed red lines show the linear fit (asymmetric barrier fit: $p$-value $= 8.87 \times 10^{-6}$, $R^{2}=0.12$, symmetric barrier fit: $p$-value $ = 0.004$, $R^{2}= 0.57$) and the confidence bounds, respectively. }

\label{SI_figure_7}
\end{figure*}

\begin{figure*}%[hbtp]
\centering
\includegraphics[width=1\linewidth ]{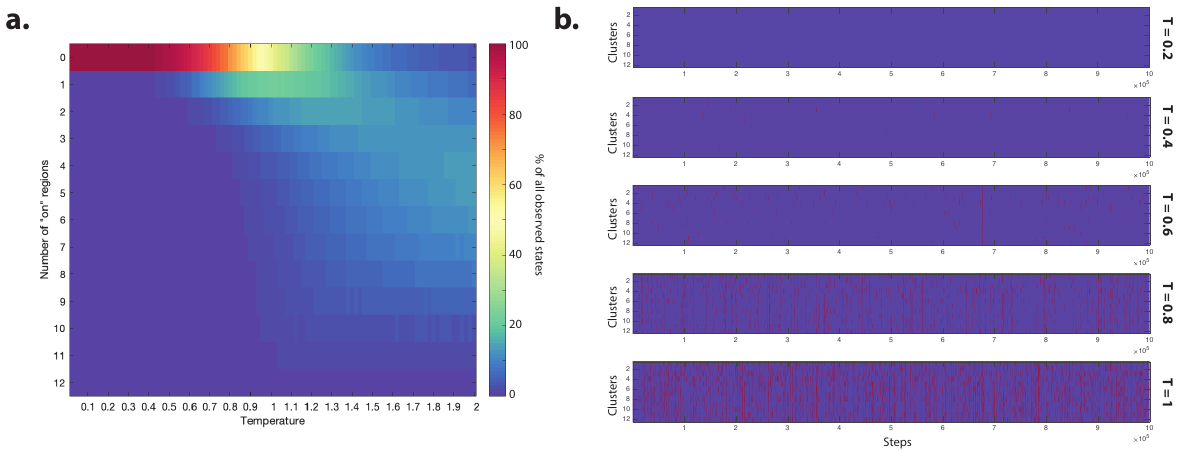}
\caption{\textbf{Simulated phase transition of the pairwise MEM at higher binarization threshold of cluster activity ($z=1$).} \textbf{a.} The plot shows the percentage of simulated states with different numbers of "on" clusters at different temperatures, using the Monte Carlo Markov Chain (MCMC) simulation method. \textbf{b.} MCMC-simulated state transitions at five different sample temperatures. The "on" clusters are highlighted in red.  Note that in the Ising model of the ferromagnetism, the spins' direction in the absence of an external magnetic field (i.e., $h = 0$) is randomly assigned, the phase transition to ferromagnetic is symmetric as the net magnetic moment can point to either direction. However, the presence of an external magnetic field (i.e., $h \neq 0$) can lead to symmetry breaking. For instance, our simulations show that the overall negative distribution of the estimated $h$ (i.e., clusters' activation propensity) promotes transition to the silent state following cooling (i.e., increased global connectivity) at higher cluster activation binarization thresholds (e.g., $z=1$). However, the brain-wide active state's duration depends on the average cluster activation time series binarization threshold. Lowering the binarization thresholds results in more prolonged global active and seizure-like states at lower temperatures (e.g., $z=0$ in manuscript Fig. 4).} 
\label{SI_figure_8}
\end{figure*}

\begin{figure*}%[hbtp]
\centering
\includegraphics[width=.7\linewidth ]{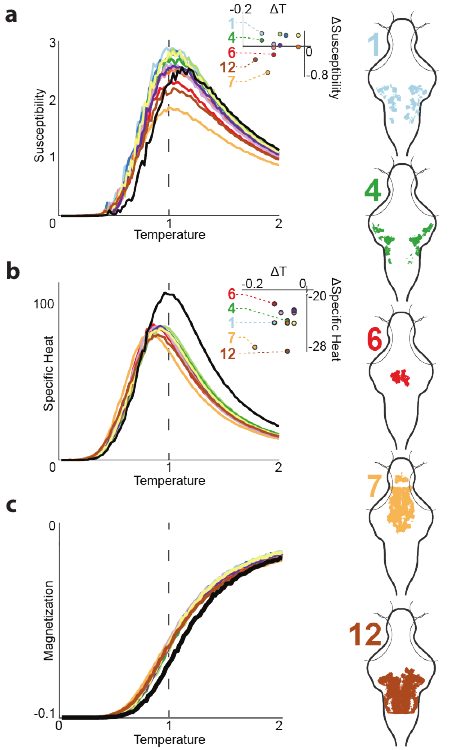}
\caption{\textbf{Phase transitions of pairwise MEM of the zebrafish brain at higher binarization threshold of cluster activity ($z=1$).}  \textbf{a.} The susceptibility,  \textbf{b.} specific heat \textbf{c.} and magnetization curves before (black) and after (color-coded for each cluster) virtual resection. The insets show the change in the peak of the curves following the resections.} 
\label{SI_figure_8}
\end{figure*}

% \begin{figure*}%[hbtp]
% \centering
% \includegraphics[width=0.6\linewidth ]{Node_Strength_Vs_Delta_Tc_12_MEM_MaxLikelihood_Thresh_1.png}
% \caption{\textbf{The correlation between the change in critical temperature following the virtual resection of the cluster and the strength of its functional connections among clusters.} Each dot represents a different functional cluster and displays the post-resection change in critical temperature (measured from specific heat curves) on the y-axis and the strength of the cluster's functional connections (i.e., the sum of elements on the corresponding $J_{ij}$ row) on the x-axis. The solid and dashed lines show the linear fit and confidence bounds ($p-$value$=0.02$, $R^{2} = 0.74$).   }
% \label{SI_figure_9}
% \end{figure*}

% \begin{figure*}%[hbtp]
% \centering
% \includegraphics[width=0.6\linewidth ]{Cluster_Remove_Weighted_FWHM_12_MEM_Thresh_1.png}
% \caption{\textbf{Resection-induced changes in heat curve width.} The full width at half maximum (FWHM) values of the post-resection curves for 12 clusters are presented, with each curve's FWHM value divided by its peak weight. A dashed black line indicates pre-resection FWHM.}
% \label{SI_figure_10}
% \end{figure*}

\newpage

\bibliography{journalBiblioSupp}
\bibliographystyle{naturemag}